# Filling a gap in materials mechanics: Nanoindentation at high constant strain rates upto $10^5$ $s^{-1}$

Lalith Kumar Bhaskar*[1], Dipali Sonawane[1], Hendrik Holz[1], Jeongin Paeng[1], Peter Schweizer[1], Jing Rao[1], Bárbara Bellón[1], Damian Frey[2], Aloshious Lambai[3,4], Laszlo Pethö[5], Johann Michler[5,6], Jakob Schwiedrzik[7], Gaurav Mohanty[3], Gerhard Dehm*[1], Rajaprakash Ramachandramoorthy*[1]

[1] Max-Planck-Institute for Sustainable Materials, Department of Structure and Micro-/Nano- Mechanics of Materials, Max Planck-Strasse 1, 40237 Düsseldorf, Germany

[2] Alemnis AG, Schorenstrasse 39, 3645 Thun, Switzerland

[3] Materials Science and Environmental Engineering, Faculty of Engineering and Natural Sciences, Tampere University, 33014 Tampere, Finland

[4] Current address: VTT Technical Research Centre of Finland Ltd., Kemistintie 3, FI-02044 Espoo, Finland

[5] Laboratory of Mechanics of Materials and Nanostructures, Empa – Swiss Federal Laboratories for Materials Science and Technology, Feuerwerkerstrasse 39, 3602 Thun, Switzerland

[6] Ecole Polytechnique Fédérale de Lausanne, Institute of Materials (IMX), CH-1015 Lausanne, Switzerland

[7] Laboratory for High Performance Ceramics, Empa – Swiss Federal Laboratories for Materials Science and Technology, Ueberlandstrasse 129, 8600 Dübendorf, Switzerland

*Corresponding author: l.bhaskar@mpie.de; g.dehm@mpie.de; r.ram@mpie.de

## Abstract

A central focus in high strain rate research is understanding the dynamic behaviour of materials at strain rates where a strength upturn is observed. While strength upturns at strain rates of $10^3$ to $10^4$ $s^{-1}$ have been widely reported in the literature, their occurrence in certain materials remains controversial, and the underlying physics driving this phenomenon is not yet fully understood. Current mechanical testing methods are limited, as no single technique spans the full strain rate range of $10^1$ to $10^5$ $s^{-1}$ where this phenomenon is expected, and a unified technique would enable consistent post-deformation characterization with minimal error. To address this, we developed a customized piezoelectric *in situ* nanomechanical test setup, enabling constant indentation strain rates up to $10^5$ $s^{-1}$, for the first time. Using this system, we examined rate-dependent hardness in single-crystalline molybdenum, nanocrystalline nickel, and amorphous fused silica over strain rates from $10^1$ to $10^5$ $s^{-1}$, remarkably revealing a hardness upturn in all three materials. Further, post-deformation analysis of single-crystalline molybdenum revealed that the hardness upturn was primarily driven by increased dislocation density, with phonon drag—traditionally considered a dominant contributor—playing a minimal role.

## Introduction

Human civilization is currently in the midst of the Fourth Industrial Revolution, marked by significant advancements in technologies such as high-speed machining, space vehicles (micro-asteroid impacts), cold spray, and micro/nanoelectronics (MEMS/NEMS) where materials are exposed to extreme demands, including high strain rates[1,2]. Despite these technological advancements, characterizing material behaviour at high strain rates remains a challenge, as much of the current testing methodologies, especially at the small scales, are still based on quasi-static approaches. A research gap in the nano-to-meso scales persists in reliable quantitative mechanical testing methodologies, particularly for strain rates between $10^2$ and $10^5$ $s^{-1}$. In the macroscale, efforts to study dynamic material behaviour have led to the development of techniques like Kolsky (Split-Hopkinson) bars, gas-gun driven projectiles, shock impacts, and others[3–5]. These macroscopic methods, owing to the large sample size, require extreme impact velocities (upto even a few km/s) to generate high strain rates, typically providing reliable data up to $10^4$ $s^{-1}$. Beyond this strain rate, shock effects complicate material response interpretation [6]. In terms of microscale testing, recent micro-ballistic advancements such as the laser-induced particle (~2-40µm in diameter) impact test (LIPIT) allow material studies at even higher strain rates ($10^6$ to $10^9$ $s^{-1}$)[7–10], providing insights into dynamic hardness and corresponding deformation mechanisms at such extreme strain rates[8,9,11]. However, it must be pointed out that quantitative load-displacement curves cannot be obtained by this technique. In addition, recent efforts have led to the development of miniature Kolsky bar systems to test micro-scaled pillars[12,13], enabling testing at higher strain rates in the range of $\sim 4 \times 10^3$ to $10^6$ $s^{-1}$. However, these systems operate just above the critical strain rate regime where the strength upturn is commonly observed, leaving a gap in detailed mechanistic understanding of how materials respond in this transitional range of strain rates.

In terms of material physics, macroscale high strain rate experiments have revealed that many materials show increased strain rate sensitivity, or the strength upturn phenomenon[14–17], at strain rates between $10^3$ and $10^4$ $s^{-1}$. In metals, this is attributed to a shift from thermally activated dislocation motion at lower strain rates to rate-dependent dislocation structure evolution and/or dislocation-phonon drag interactions at higher rates[18–20]. While the upturn phenomenon is widely accepted, there are still ongoing debates about the exact strain rate at which it occurs and the deformation mechanism responsible. Some investigations from macroscale experiments, report the upturn between $10^3$ and $10^4$ $s^{-1}$ strain rate,[15,21,22] while others suggest that data scatter and experimental artifacts (such as elastic wave propagation) at these rates are responsible for the observed effect[12,23]. A comprehensive review of this controversy in macroscale high strain rate experiments can be found in the work of Z. Rosenberg *et al*[24]. Conversely, the microscale LIPIT technique enables high strain rate explorations only higher than $10^6$ $s^{-1}$, creating a gap in strain rate dependent testing of materials. Both dynamic macroscale and LIPIT microscale techniques allow the high strain rate investigation of materials but cannot typically maintain constant strain rates during the entire duration of the experiment, further complicating post-deformation analysis to understand the underlying deformation mechanisms.

Instrumented nanoindentation offers a promising solution for high strain rate testing, enabling constant strain rate experiments at the micron and sub-micron scales in a variety of material systems while uniquely allowing testing of small phases or features inaccessible to other methods[25,26]. While traditionally limited to quasi-static strain rates below $10^{-1}$ $s^{-1}$, recent efforts have extended nanoindentation to higher rates. Impact-based techniques, using pendulum actuators[27–29], can achieve an average high strain rate of $10^4$ $s^{-1}$ but lack precise control over load, displacement, and strain rate, complicating deformation analysis. Advanced nanomechanical systems now allow hardness measurements across strain rates of $10^1$ to ~$10^4$ $s^{-1}$, but these methods face challenges, including the inability to maintain a constant strain rate beyond $10^2$ $s^{-1}$ and difficulties in determining initial step loads for achieving these strain rates, often requiring complex numerical solutions[30–34]. Displacement-controlled systems [35–37] have achieved constant indentation strain rates up to $10^3$ $s^{-1}$, but this remains an order of magnitude below the rates where an upturn in material strength is expected, underscoring the need for further advancements in high constant strain rate nanoindentation techniques.

In this study, we present a piezoelectric-based in situ high-speed nanomechanical testing setup capable of achieving constant indentation strain rates ($\dot{h}/h$), for the first time, from $10^1$ $s^{-1}$ upto $10^5$ $s^{-1}$. Such extreme testing also necessitated the development of highly custom-modified support electronics and experimental protocols for capturing precise load-displacement data during these high-strain-rate indentations. Further, new methodologies were also devised for extracting accurate hardness values from the load-displacement curves captured at high strain rates. This enabled the assessment of hardness and deformation behaviour in single-crystalline molybdenum, nanocrystalline nickel, and amorphous fused silica at unprecedented constant indentation strain rates. Comprehensive post-indentation analyses were performed using confocal laser microscopy, quasi-static reloading, and micro-pillar compression for fused silica to further validate the observed hardness trends. Moreover, transmission electron microscopy (TEM) of molybdenum, in conjunction with theoretical calculations, provided new insights into previously unresolved questions in fundamental material science governing deformation behavior at high strain rates. By enhancing our understanding of deformation mechanisms at extreme strain rates, these findings open new avenues for tailoring the mechanical properties of materials through strain-rate-dependent microstructural engineering.

## High constant strain rate nanoindentation and upturn in hardness

This study presents, for the first time, high constant strain rate nanoindentation ($\dot{h}/h$) from $10^1$ $s^{-1}$ to $10^5$ $s^{-1}$ using a modified *in situ* nanomechanical testing platform (Alemnis AG) as shown in Fig. 1A. Detailed instrumentation modifications and methods for capturing reliable load and displacement data over short time spans (~150 μs) are outlined in the Methods section. A significant breakthrough in this work is the ability to maintain a constant indentation strain rate, which enabled accurate investigation of the underlying deformation mechanisms and their correlation with specific strain rates (up to $5\times10^4$ $s^{-1}$). While the stress state beneath the indenter tip is inherently non-homogeneous, post-deformation microstructural analysis was carefully conducted in a consistent spatial region beneath the indent across all strain rates. This ensured that variations in the observed microstructure could be reliably attributed to differences in strain rate, rather than variations in stress localization or sampling position. Although higher indentation depths were tested, the strain rate plot in Fig.1B is limited to 700 nm, beyond which the strain rate drops at the highest tested strain rate of $10^5$ $s^{-1}$, as shown in fig. S1. This drop in strain rate beyond 700nm indentation depth is due to the inability of the high-speed, high-voltage amplifier to supply an amplified voltage output with extreme voltage gradients (slew rate > 75 V/μs) needed to maintain a constant strain rate of $10^5$ $s^{-1}$. To maintain consistency, all further results and analyses presented here are limited to 700 nm depth across all strain rates.

The corrected load-displacement curves for single crystalline molybdenum obtained after applying all protocols outlined in the Methods section are shown in Fig. 2A, while those for nanocrystalline nickel and amorphous fused silica are presented in fig. S2. In the load-displacement curves, only the loading portions are shown, as resonance effects from various components of the testing system significantly affect the unloading segments, as shown in fig. S3. Consequently, the traditional Oliver-Pharr method[25] for hardness determination was inapplicable. Instead, two other complementary techniques, the Merle-Higgins-Pharr (MHP iterative method) [38] and projected area methods, were used (see Methods section). Fig. 2B–D present rate-dependent hardness values for molybdenum, nanocrystalline nickel, and fused silica obtained from both methods, compared against literature values. Fig. S4A-C shows hardness versus displacement for all three materials using the MHP iterative method.

Using the MHP iterative method, hardness values were determined across the full displacement range; thereafter, the Nix-Gao model[39] was applied to fit data between 200–700 nm for molybdenum and nanocrystalline nickel, where strain rates remained constant for all strain rates, before being extrapolated for direct comparison with hardness values obtained using the projected area method. In contrast, the projected area method provides a single hardness data point at the highest load. Fig. S5A indicates that at the highest load, the strain rate is lower than the expected strain rate. Consequently, the hardness values from the projected area method in Fig. 2B–D were adjusted to correspond to the strain rate ($\dot{P}/2P$ – from load), calculated by averaging over a range of 10–20 nm at the peak load. In this work, for hardness calculations using the projected area method, the peak load after all corrections was used (fig. S5B). Although nanocrystalline nickel is generally not expected to exhibit an indentation size effect[40], the Nix–Gao model was nonetheless employed for this material as well. The rationale for this choice is detailed in Supplementary Information – *Section S1*. However, for fused silica, no Nix–Gao correction was applied since it does not exhibit indentation size effects. Instead, hardness values were averaged over a depth range of 600–650 nm. Additionally, Fig. 2D presents hardness measurements for fused silica at strain rates of $5\times10^{-2}$ $s^{-1}$ and $10^{0}$ $s^{-1}$, obtained using two independent systems – KLA G200 and the quasi-static system from Alemnis AG[36], respectively. The observed hardness trends are consistent and align well with those obtained using the modified high strain rate indentation setup.

In Fig. 2B, the hardness trends for molybdenum show a strong correlation between the two complementary methods, with a distinct upturn in hardness observed at strain rates exceeding $3\times10^{3}$ $s^{-1}$. Beyond this strain rate, the calculated strain rate sensitivity (given by, $m = \frac{\partial lnH}{\partial ln\dot{\varepsilon}}$, where H is hardness and $\dot{\varepsilon}$ is strain rate) increases approximately- 2.5 times from m = 0.044 $\pm$ 0.009 to 0.112 $\pm$ 0.011. Previous high strain rate impact nanoindentation study up to $10^{4}$ $s^{-1}$ on molybdenum[30] with (110) orientation showed a similar value of strain rate sensitivity (m = 0.032 $\pm$ 0.002). However, no hardness upturn was reported in this study for the (100) orientation. This variation could be attributed to differences in the sample orientation and the testing methodology used in the literature, where experiments resulted in varying strain rates and instantaneous strain rates were used to report hardness values [30].

Similarly, for nanocrystalline nickel in Fig. 2C, the hardness values from both techniques showed good agreement, with an upturn observed at strain rates above $10^3$ $s^{-1}$ and an approximately four-fold increase in strain rate sensitivity (from m = 0.029 $\pm$ 0.013 to 0.115 $\pm$ 0.008). Literature reports of constant strain rate indentation up to $10^3$ $s^{-1}$ showed similar rate sensitivity (m = 0.022 $\pm$ 0.009)[36]. Notably, the upturn in hardness observed in this work aligns with dynamic compression experiments using Kolsky bars[41], which also reported a strength increase above $10^3$ $s^{-1}$. While the grain sizes in this study (38 nm) are comparable to those reported in the literature (17 nm [41] and 26 nm [36]), the variations in absolute hardness values in Fig. 2C are likely due to chemical impurities introduced by electrodeposition conditions, which are known to affect the microstructure and, consequently, the mechanical properties[42].

For fused silica, unlike molybdenum and nanocrystalline nickel, the hardness and strain rate sensitivity values from the two complementary techniques showed differences from $10^3$ $s^{-1}$ and above as shown in Fig. 2D. Interestingly, the iterative method did not provide any solutions for hardness above $10^3$ $s^{-1}$. The strain rate sensitivity (m) was calculated as 0.009 $\pm$ 0.001 up to strain rates of $3\times10^3$ $s^{-1}$, consistent with literature[43]. Beyond this threshold, however, an upturn in hardness was observed, with a strain rate sensitivity of 0.064 $\pm$ 0.004 — a 7-fold increase.

**Analysis of the deformation mechanism behind hardness upturn**

The literature data[33] for fused silica also shows a similar upturn in hardness. However, authors in that study expressed caution, suggesting it might be a data processing artifact caused by filtering rather than a genuine material response. There are two key questions to be answered for fused silica: (i) why did the iterative method fail to provide a hardness solution beyond $10^3$ $s^{-1}$, and (ii) is the observed upturn in hardness a true material behaviour?

Fused silica, characterized by its high free volume due to low atomic packing density, predominantly deforms through densification[44]. Upon sufficient densification, it also exhibits conservative shear flow as a plastic deformation mode[45]. Literature reports indicate a maximum densification of 21%, with changes in elastic modulus becoming noticeable once densification exceeds 5%[45]. Interestingly, in this study, from micropillar compression experiments pristine fused silica pillars exhibited an increase in apparent loading modulus from 71.4 $\pm$ 0.9 GPa to 83.4 $\pm$ 0.4 GPa under compression at strain rates exceeding ~$10^3$ $s^{-1}$ as shown in Fig. 3A & 3B. This is hypothesized to occur because, at high strain rates, the intense hydrostatic forces generated by the taper of the fused silica micropillars may induce reversible instantaneous densification, potentially leading to a change in elastic modulus. Although this hypothesis aligns with literature evidence that supports an increase in elastic modulus with densification[45,46], the direct experimental evidence of densification remains elusive as it occurs within the elastic region. However, similar trends in loading modulus have been reported in other compression studies on fused silica at comparable strain rates [35].

Further to rule out the possibility of a machine-induced artifact causing the observed modulus change, compression tests were conducted on fused silica pillars of two different heights—5.65 µm and 14.5 µm. Notably, both pillar geometries showed a change in loading modulus at the same transition strain rate of ~$10^3$ $s^{-1}$ (Fig. 3B). At this strain rate, the corresponding compression velocities were 5.65 mm/s for the 5.65 µm pillars and 14.5 mm/s for the 14.5 µm pillars. Furthermore, the hardness upturn observed in indentation experiments (Fig. 2D) aligns with the flow stress upturn seen in the micropillar compression results at similar strain rates (Fig. 3B). Specifically, this transition corresponds to an instantaneous indentation velocity of just 0.7 mm/s for a 700 nm indentation depth, reinforcing that the observed upturns are consistent across different testing methods on the same material, thereby further ruling out machine-related artifacts. The stress-strain curves for the 5.65 µm high pillars are given in Fig. 3A, and those of the 14.5 µm high pillars are shown in the supplementary fig. S6.

The observed modulus changes in fused silica micropillars beyond ~$10^3$ $s^{-1}$ can explain why the iterative hardness estimation method fails, as it relies on the assumption of a constant reduced modulus across strain rates (see Methods section). However, the reason for the observed upturn in nanoindentation hardness, while potentially attributable to modulus changes, remains speculative and could also result from a shift in deformation mechanisms, as the depth profiles of confocal microscopy images of fused silica across strain rates (fig. S7A) reveal small pile-ups at lower strain rates, indicating a potential shear flow mechanism[45], whereas these pile-ups are absent at higher strain rates. These trends are consistent with previous findings in glassy materials, where high strain rate-induced transitions from shear band propagation to shear band nucleation[47], or alternatively, structural responses governed by the time scale of crack propagation in brittle materials[48], have been proposed. Despite these compelling parallels, the underlying mechanism remains unresolved. Significantly, even the loading modulus in this study exhibits clear strain-rate sensitivity, highlighting the need for further investigation into the high strain rate behaviour of fused silica using advanced characterization techniques such as synchrotron-based ptychographic X-ray computed tomography (PXCT)[49], coupled with atomistic based continuum-scale simulations.

Unlike fused silica, the line profiles for molybdenum and nanocrystalline nickel showed no significant difference in pile up across strain rates, as illustrated in fig. S7B & S7C. The observed upturn in hardness in these metallic materials should correlate with changes in the dislocation substructure or microstructure beneath the indent. To confirm this, the same indents formed during high strain rate indentation were reloaded[30] at a lower strain rate of $10^1$ $s^{-1}$. Details of this reloading process are provided in the Methods section, and a simplified schematic of the procedure is shown in Fig. 3C. The hardness obtained during reloading, along with the hardness trend as a function of strain rate for molybdenum and nanocrystalline nickel, is presented in Fig. 3D and 3E, along with the load-displacement curves in fig. S8A & S8B. Notably, the strain rate at which the upturn in hardness occurs coincides with a distinct increase in reloading hardness, indicating a possibility of microstructural/dislocation substructure change. Similar trends were observed in the reloading hardness of fused silica, as shown in fig. S8C.

**Rate dependent microstructure evolution in metals**

To better understand the rate-dependent deformation mechanism in metals, a combination of thermal activation analysis, microstructural characterization via TEM, and theoretical calculations was conducted. The apparent activation volume ($\Omega_{app}$) was calculated and expressed in terms of the Burgers vector ($b$), as shown in Fig. 2B and 2C. Before the hardness upturn, for molybdenum, $\Omega_{app}$ was $2.47 \pm 0.43\ b^3$, consistent with literature values at room temperature [50], indicating the formation and migration of dislocation kink pairs typical for BCC metals. For nanocrystalline nickel, $\Omega_{app}$ was $1.94 \pm 0.89\ b^3$, reflecting grain-boundary mediated deformation typical for FCC nanomaterials, aligning well with reported values [40,51]. However, beyond the hardness upturn, the apparent activation volumes decreased significantly, with molybdenum and nanocrystalline nickel showing $0.74 \pm 0.06\ b^3$ and $0.43 \pm 0.03\ b^3$ respectively. Interestingly, such low activation volumes suggest the rate dependence is due to dislocation nucleation events [52,53].

At low strain rates, thermally activated dislocation motion is the primary deformation mechanism. In this regime, dislocations overcome atomic obstacles with the assistance of thermal vibrations. However, as strain rates increase, thermal activation becomes less effective, requiring higher stresses for dislocation glide. This limitation arises because, for a given dislocation density ($\rho$), there is a maximum dislocation velocity ($\nu_{max}$), which sets an upper limit on the plastic strain rate ($\dot{\varepsilon}^p_{max} = \rho b \nu_{max}$) as per the Orowan equation. It was observed from Discrete dislocation dynamics (DDD) simulations that if the applied strain rate surpasses this threshold, the existing dislocations beneath the indent become insufficient to accommodate the imposed stress, resulting in a substantial increase in dislocation nucleation— a phenomenon referred to as the exhaustion zone [18]. Similar theories were proposed by Follansbee et al. [15], based on macroscale high-strain-rate experiments. However, obtaining direct evidence of an increase in dislocation density remained challenging, as only an approximately 1.5 fold increase was expected [15], and earlier attempts using transmission electron microscopy were not successful in detecting significant changes in dislocation densities accurately[54]. Furthermore, recent constitutive modeling by Tang et al. suggests that dislocation-phonon drag, a prominent mechanism considered responsible for the upturn in strength, becomes significant only at strain rates exceeding $10^5$ $s^{-1}$, whereas the upturn observed at $10^3$ $s^{-1}$ is primarily attributed to thermal activation effects [14].

To investigate these claims, a detailed post-deformation microstructural analysis of indents in molybdenum was performed using TEM. Molybdenum was selected due to its single-crystalline structure, which simplifies interpretation compared to nanocrystalline nickel. To identify the underlying deformation mechanisms, TEM analysis was done on the indents conducted at the lowest tested strain rate $10^1$ $s^{-1}$, just before the upturn $3\times10^3$ $s^{-1}$, and at the highest strain rate $5\times10^4$ $s^{-1}$ at which the strain rate remains constant for the majority of the indentation depth. Fig. 4 (A-C) presents annular bright field scanning transmission electron microscopy (ABF-STEM) images of indents at three different strain rates, revealing clear qualitative differences in dislocation density. The dislocation densities were also quantified, with details provided in the Methods section.

The measured values are $(9.1 \pm 2.4)\times10^{13}$ $m^{-2}$, $(1.3\pm0.3)\times10^{14}$ $m^{-2}$, and $(3.2\pm0.9)\times10^{14}$ $m^{-2}$ at strain rates of $10^1$ $s^{-1}$, $3\times10^3$ $s^{-1}$, and $5\times10^4$ $s^{-1}$ respectively, showing an approximately 3.5-fold increase in dislocation density across strain rates. Notably, scanning electron microscopy (SEM) images of the indent cross-sections in Fig. 4 (A-C) indicate that the indentation depth remains nearly constant across all three strain rates. Dislocation densities were notably measured at a consistent distance of approximately 1.5 μm from the indent apex across all strain rates. This depth was chosen for ease of measurement and to confirm the relative trends across different strain rates. However, it is to be noted that dislocation densities would be substantially higher near the indenter tip. However, this study represents the first instance where dislocation density at such high strain rates has been imaged and comparatively quantified across strain rates with absolute certainty.

With the dislocation density values determined, we used them to investigate the possible deformation mechanism responsible for the observed upturn in hardness. Since the material under consideration is single crystalline, the possible strengthening mechanisms are limited to lattice resistance strengthening, dislocation strengthening, and dislocation drag. Detailed calculations for each of these mechanisms are provided in Supplementary Information – *Section S2*, and the resulting trends are plotted in Fig. 5 across these three strain rates. For comparison, the measured hardness values were normalized by dividing by a constraint factor of 2.8 and plotted alongside the calculated strengths. Fig. 5 clearly shows that dislocation strengthening is the dominant mechanism driving the upturn in strength. Specifically, at a strain rate of $5\times10^{4}$ $s^{-1}$, the strength attributed to dislocation hardening is approximately 1.6 times greater than that at $3\times10^{3}$ $s^{-1}$. Additionally, Fig. 5 shows that dislocation drag, previously hypothesized in the literature as a significant contributor to the upturn in strength, has a negligible impact possibly owing to the high dislocation density. Even at strain rates as high as $5\times10^{4}$ $s^{-1}$, its contribution is minimal, with a value of only 0.024 GPa.

**Summary and outlook**

This study presents the development of a specialized high-speed piezoelectric-based micromechanical testing setup for constant strain rate nanoindentation, achieving unprecedented strain rates from $10^{1}$ $s^{-1}$ to $10^{5}$ $s^{-1}$. Custom electronics and advanced protocols ensured the acquisition of accurate load-displacement signals, enabling reliable hardness measurements using two complementary methods. The rate-dependent hardness trends of single-crystalline BCC molybdenum, FCC nanocrystalline nickel and amorphous fused silica were investigated. Surprisingly, all materials demonstrated a hardness upturn with increasing strain rate. The hardness upturn was validated through a combination of micropillar compression tests in fused silica and reloading experiments for all materials. The constant strain rate capability allowed a precise correlation of microstructural evolution underneath the indent to the hardness upturn. TEM studies at strain rates exceeding the upturn threshold revealed an increased dislocation density due to dislocation nucleation, offering direct evidence of strain rate-dependent strengthening mechanisms in single-crystalline molybdenum with negligible contributions from dislocation-phonon drag. While the exact rate-dependent microstructural evolution for nanocrystalline nickel and fused silica still need to be explored, the possibility of conducting high constant strain rate nanoindentation at a wide range of strain rates with reliable load-displacement curves, hardness data, and rate-specific microstructural investigations is a new pathway for conducting more accurate and quantitative dynamic material property investigations. Additionally, nanoindentation at such extreme strain rates on individual phases, grains, and grain boundaries, would provide valuable rate-dependent input for simulations like DDD and enable the development of more accurate physics-based models in the future [18] for several materials.

## Materials and Methods

The molybdenum sample, with 99.999% purity and a (100) crystallographic orientation, was commercially sourced from Goodfellow Cambridge Limited. It was progressively polished resulting in a mirror-like surface finish. The nanocrystalline nickel used in this study was electrodeposited in a UV-LIGA mold on a silicon wafer using a nickel sulfamate bath. The top surface was polished, and samples were subsequently extracted from LIGA molds. Optical-grade amorphous fused silica was obtained from MTI Corporation. Furthermore, in this work compression tests were performed on lithography made fused silica micropillars, which had dimensions of approximately 2.5 μm in diameter and 5.65 μm in height, and 5.5 μm in diameter and 14.5 μm in height . The detailed fabrication procedure for fused silica micropillars is described in the work of R. Ramachandramoorthy *et al.*[35]

*Ultra-high strain rate nanomechanical testing system*

The system used in this study is based on a platform from Alemnis AG, which was modified to further extend the range for high constant strain rate ($\dot{h}/h$) nanoindentation from $10^4$ $s^{-1}$ to $10^5$ $s^{-1}$. A schematic of the modified system is provided in fig. S9A & S9B. The system includes a piezoelectric load cell (Alemnis AG) for load measurements, providing a resolution of approximately 15 μN. The piezoelectric load cell is designed with four electrodes positioned on a concentric cylinder, and the signals from each electrode are collected individually. The signals are then amplified via a high-impedance charge amplifier (Alemnis AG) with a cut-off frequency of ~300 kHz. For actuation, a piezoelectric tube actuator (Alemnis AG) was employed to achieve constant indentation strain rates ($\dot{h}/h$) upto $10^5$ $s^{-1}$. The piezoelectric tube actuator features a low capacitance of 14.5 nF and a high stiffness of 19.14 N/μm. The piezoelectric tube actuator can reach very high velocities of ~ 100 mm/s, with a maximum displacement of 4 μm. To attain such high actuator velocities, an external high-speed high-voltage amplifier (WMA-200) from Falco Systems was used. The voltage amplifier can deliver output voltages of ±175 V, a maximum current output of 150mA, and has a high slew rate of 80 V/μs (no load), which is crucial for maintaining such high constant strain rates. The voltage amplifier is capable of receiving shaped input profiles with a maximum amplitude of ±10 V, which are then amplified 20-fold for output. Additionally, the voltage amplifier is equipped with a monitoring port that continuously records the amplified voltage albeit reduced by 10 times in magnitude, which was critical for the synchronization of different load and displacement signals, as detailed further in this section. To measure the actuator's displacement during indentation testing, two piezo-resistive strain gauges were attached on opposite sides of the piezoelectric tube, and a full Wheatstone bridge was formed using an external completion circuit. This was further connected to a strain gauge amplifier with a high bandwidth of ~2.5MHz. In addition to the piezo-resistive strain gauges used for displacement measurement, a laser interferometer was employed in this work to validate displacements measured via the strain gages. The laser interferometer, sourced from SmarAct Metrology (model V2 system), offers a displacement resolution of 1 pm and a measurement bandwidth of 2.5 MHz. A schematic of the displacement measurement setup is shown in fig. S9B. In this configuration, the sample is mounted on the piezoelectric tube actuator, while a laser with a wavelength of 1545 ± 15 nm and a fixed focal length of 10 ± 0.5 mm is positioned opposite the actuator. The mirror-polished surfaces of the samples are sufficient enough to reflect the laser and facilitate accurate displacement measurements. Predefined voltage profiles, corresponding to specific constant indentation strain rates, were designed and employed for recording the displacement using the laser interferometer.

For experiments conducted at the highest strain rate, the duration for the entire loading and unloading segment is approximately 150 μs, necessitating a high data acquisition rate to capture non-aliased signals. To achieve this, Rohde & Schwarz RTA 4000 oscilloscopes with a 5 GSamples/s acquisition rate were employed. During all the experiments, across various strain rates, a minimum of 50,000 data points were recorded for every test to ensure high data fidelity. The oscilloscope features four input channels and an additional trigger port for initiating data acquisition. The monitor signal from the high-voltage amplifier was used as the trigger for all experiments. To smoothen the data, all acquired signals were processed using zero-phase low-pass filters prior to any further analysis. All data processing was implemented in a custom-written program developed in LabVIEW.

*Methodology for signal acquisition and synchronization*

Predefined shaped input exponential profiles for indentation were generated by the controller and sent to a high-speed high voltage amplifier, which amplifies the signal 20-fold to drive the piezoelectric tube actuator. The sample, mounted on the piezoelectric tube actuator, was then brought into contact with the indenter tip (Berkovich or flat punch) attached to the piezoelectric load cell, to perform either indentation or compression experiments.

The displacement was measured using piezo-resistive strain gauges attached to the piezoelectric tube actuator, and the load was recorded via the charges output from the piezoelectric load cell. Both displacement and load signals were fed into a strain gauge amplifier and a high-impedance charge amplifier, respectively, and the amplified analog output signals were captured using the oscilloscope. Additionally, the amplified monitor voltage (output voltage/10) from the high-speed high-voltage amplifier was recorded during all experiments. This monitored voltage captured using the oscilloscopes served two functions: triggering the oscilloscope to begin data acquisition and aligning the signals from all sensors.

In total, seven signals were recorded during the *in situ* experiments: four electrodes of the piezoelectric load cell, one from the piezo-resistive strain gauge, and two mirrored monitored voltage signals from the high-speed amplifier (later used for synchronization using a custom written LabVIEW software). Due to the oscilloscope's four-input port limit, two oscilloscopes were used to capture all signals. The detailed schematic of the test setup is shown in fig. S9.

In this work, in addition to the piezo-resistive strain gauges, displacement was also measured using a laser interferometer, with the experiments being conducted *ex situ* on an optical table. The use of a second displacement measurement system was necessitated by the observation that the strain gauges mounted on the actuator provided inaccurate displacement readings when compared to those from the laser interferometer at very high strain rates ($10^5$ $s^{-1}$), as demonstrated in fig. S10A. Several factors may contribute to the underperformance of strain gauges at such high velocities, including the mounting configuration of the strain gauges (strain gauges measure locally whereas laser interferometer directly measures at sample position) and the amplification errors/oscillations from the operational amplifiers [55]. It is important to note, however, that the piezo-resistive strain gauges perform reliably, with deviations only observed in the last 20 nm of displacement at strain rates up to $10^4$ $s^{-1}$, as shown in fig. S11. The signal quality from the strain gauges deteriorates only beyond this strain rate. Although the laser interferometer proved highly reliable for displacement measurements, it could not be incorporated into the *in situ* experimental setup due to the compact design of the micromechanical test system. Consequently, displacement measurements using the laser interferometer were conducted *ex situ*, as illustrated in fig. S9B, where the lower portion of the micromechanical setup, including the actuator and sample, was mounted on an optical table for displacement recording. To maintain consistency across all strain rates ($10^1$ to $10^5$ $s^{-1}$), the displacement data recorded via the laser interferometer was used for generating all the load-displacement reported here. The same predefined voltage profiles used during the *in situ* indentation experiments were used in the *ex situ* laser interferometer displacement measurement experiments. Since the monitored voltage from the high-speed voltage amplifier was consistently recorded in the oscilloscopes, these signals were utilized to synchronize the load data from *in situ* experiments with the displacement data from *ex situ* experiments, using a custom LabVIEW based program.

It was also critical to check, whether the presence of the sample, indenter tip, and the piezoelectric load cell reduces the piezoelectric tube actuator displacement during *in situ* experiments when compared to the *ex situ* experiments without resistance from the tip and load cell. This concern was mitigated by the fact that the actuator stiffness (~$19.14 \times 10^6$ N/m) was at least two orders of magnitude higher than the sample stiffness (~$10 \times 10^4$ - $25 \times 10^4$ N/m). In other words, the micromechanical system operates as an intrinsic displacement-controlled setup. To further validate that the displacement was not reduced with or without the sample, fig. S10B compares the *ex situ* displacement measured by the laser interferometer and the *in situ* displacement recorded during an indentation experiment at $10^1$ $s^{-1}$, showing a negligible difference of $13 \pm 2$ nm.

*Time Synchronization*

Since the load and displacement signals are captured by different electronic systems (outlined in the previous section) with differing processing times, a time lag arises between the two signals. To quantify this lag, the diamond Berkovich indenter tip was manually plunged into the sample surface using the piezo based positioning stages (on all three materials of molybdenum, nanocrystalline nickel, and fused silica) until distinct load signals were detected. While keeping the tip in contact with the sample surface, a near step-like voltage profile was applied to the piezotube actuator and the signals were collected using an oscilloscope in a highly magnified temporal window such that even minimal time lags as small as 0.5 μs can be captured, as shown in the inset of fig. S12A. From this calibration, a delay of $4.98 \pm 0.05$ μs was calculated between the load and displacement signals, as illustrated in fig. S12A. The displacement data was also recorded *ex situ* using a laser interferometer, following the same voltage profile used *in situ* for load measurement. Synchronization between the *in situ* load signal and *ex situ* displacement data was achieved using the monitor voltage, aligned through a custom LabVIEW based program. The interferometer's displacement signal showed a longer lag than the load signal, attributed to the extended processing time of the interferometer's FPGA compared to the high-impedance charge amplifier used for the load cell. Consequently, the displacement signal trails the load signal, as seen in fig. S12A. Fig. S13 illustrates the resulting errors and time lag effects on load-displacement data, becoming noticeable and propagating at strain rates above $10^3$ $s^{-1}$.

*Time Constant*

For high-speed measurement systems, the time constant of a sensor is defined as the time for a sensor to reach about 63.5% (1 - 1/e) of its steady-state step input [56]. It is influenced by both the sensor's internal properties and associated electronics, including amplifiers. In systems like piezoelectric load cells, a shorter time constant enables faster response to dynamic load changes, which is essential for high strain rate experiments where measurements must occur within microseconds. A longer time constant, conversely, can introduce signal delays and smooth out transient events, leading to inaccuracies. To determine the time constant ($t_c$) of the piezoelectric load cell in this study, we conducted compression tests on silicon pillars (5 μm diameter, 15 μm height), which are brittle and exhibit a sharp fracture. As seen in fig. S12B, the load versus time data showed an exponential decay after fracture, rather than an instantaneous drop marked by dotted black line, allowing us to fit this data to find a time constant of 1.70 ± 0.03 μs. Fig. S14 highlights the underestimation of load at higher strain rates when the time constant is not considered, with errors appearing above $10^3$ $s^{-1}$. In this work, displacement measurements were captured using a laser interferometer with near-instantaneous recording capabilities, eliminating the need for time constant correction for displacement data. While time synchronization and time constant effects mainly impact data at strain rates above $10^3$ $s^{-1}$, corrections were applied to lower strain rate data as well to ensure consistency across all analyses. Fig. S15 displays the load-displacement curves for all materials tested in this study after applying time synchronization, time constant correction, compliance correction, and zero-displacement correction. The zero-displacement correction was determined via the Hertzian contact method to accurately establish the point of load initiation.

*Machine dynamics*

In nanoindentation experiments aimed at achieving constant strain rates, the input displacement profiles are typically designed as exponential functions of time. However, upon double differentiation of these exponential profiles with respect to time, it is clear that the acceleration as a function of depth is also not constant and it continuously varies throughout the experiment. This varying acceleration introduces challenges due to the inertial forces that arise, as dictated by Newton's second law, whereby an increase in acceleration leads to a proportional increase in inertial force (F=ma). As strain rates increase, the duration of the exponential displacement profile becomes shorter, resulting in acceleration changes over shorter periods of time. Consequently, the inertial forces acting on the components of the nanoindentation system (such as the load cell) become more pronounced at higher strain rates. Inertial forces can distort the measured load data by introducing artifacts. As shown in fig. S12C, the piezoelectric load cell and the piezoelectric tube actuator are arranged in series, with the load cell carrying its mass and movement in response to the external force applied by the actuator via the sample. Therefore, the inertial forces from the load cell must be accounted for when processing the raw load data. To calculate these forces, the system was modeled as a single-degree-of-freedom damped harmonic oscillator. The piezoelectric load cell's response to the external force (F), was modelled using a Kelvin-Voigt system. This model includes a viscous dashpot with a damping constant b=10.217 Ns/m and an elastic spring with stiffness k=32.208×10$^6$ N/m, arranged in parallel with a mass of m=0.238 g, as illustrated in fig. S12C. Details of the calculations for each parameter, along with a comprehensive methodology and error analysis for the machine dynamics corrections applied in this study, are provided in detail in Supplementary Information – *Section S3*. Fig. S12D shows the loads attributable to machine dynamics across various strain rates for molybdenum, as determined by the machine dynamics model. At the highest strain rate, machine dynamics account for a reduction of approximately 16.8 ± 1.6 μN in force at around 700 nm of displacement, equating to roughly 6% of the actual load on the sample. The dynamic load contribution varies based on the force acting on the piezoelectric load cell, and the calculated dynamic loads for fused silica and nanocrystalline nickel are presented in fig. S16. These figures indicate that the load contribution due to machine dynamics becomes significant only at strain rates 10$^4$ s$^{-1}$ and beyond for this specific system configuration. Fig. S12E provides a perspective on the cumulative effect of all three corrections—time synchronization, time constant, and machine dynamics—applied to an experiment conducted on molybdenum at a strain rate of 10$^5$ s$^{-1}$.

The detailed methodology of *nanoindentation experiments*, *micropillar compression testing, quasi-static reloading experiments, hardness measurement, and TEM, along with dislocation density measurement, is* given in Supplementary Information *Section S4.*

**Acknowledgements**

Dr. Jean-Marc Breguet from Alemnis AG is acknowledged for all the comments and technical discussions. Leon Christiansen is acknowledged for his help with SEM facilities.

**Funding**

L.K.B., J.P. and R.R. would like to acknowledge funding from the European Research Council (ERC) (Starting grant agreement No. 101078619; AMMicro). D.S. and R.R. would like to acknowledge funding from the Eurostars Project HINT (01QE2146C). D.S. and B.B. would like to acknowledge funding from the Alexander von Humboldt Foundation, and B.B., in addition, would like to acknowledge funding from the Marie Sklodowska-Curie individual fellowship (Grant agreement No. 101064660; DyThM-FCC). Views and opinions expressed are however those of the author(s) only and do not necessarily reflect those of the European

Union or the European Research Council. Neither the European Union nor the granting authority can be held responsible for them.

**Author contributions:**

Conceptualization: L.K.B, G.D. and R.R
Methodology: L.K.B and R.R
Investigation: Everyone
Funding acquisition: G.D and R.R
Supervision: R.R
Writing – original draft: L.K.B and R.R
Writing – review & editing: Everyone

**Competing interests:**

The authors declare no competing interests.

**Data and materials availability:**

All data are available in the main manuscript or the supplementary materials.

# Main figures

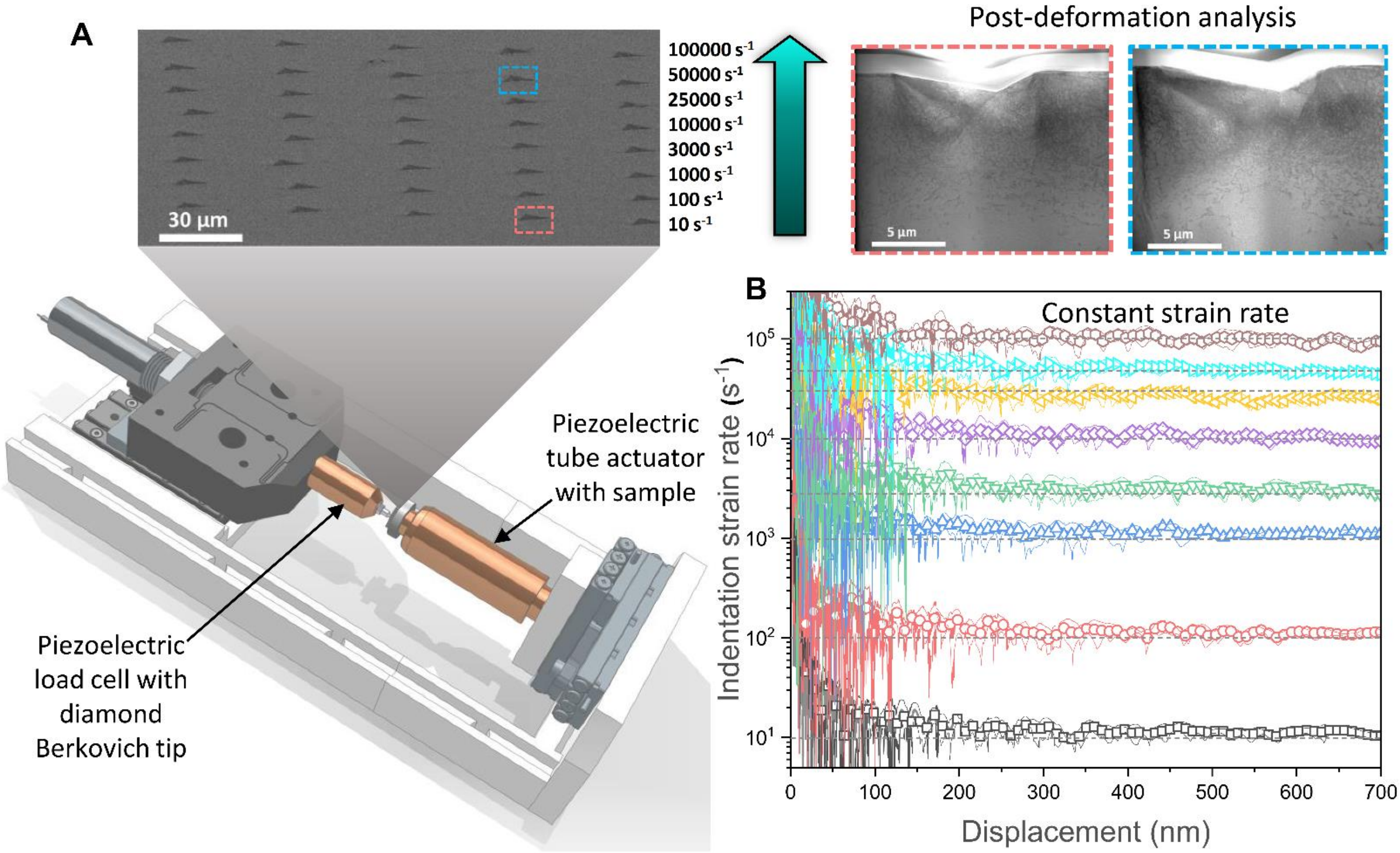

Fig.1 **High constant strain rate piezoelectric nanoindentation system**

(**A**) The schematic of the nanomechanical instrument with which nanoindentation tests were conducted over five orders of strain rate, from $10^1$ s⁻¹ to $10^5$ s⁻¹ and the representative indents on molybdenum across all strain rates. The ability to perform post-deformation microstructural analysis of the material to investigate the underlying mechanisms at each strain rate individually is additionally shown. (**B**) the constant indentation strain rate ($\dot{h}/h$) from $10^1$-$10^5$ $s^{-1}$ up to an indentation depth of 700 nm.

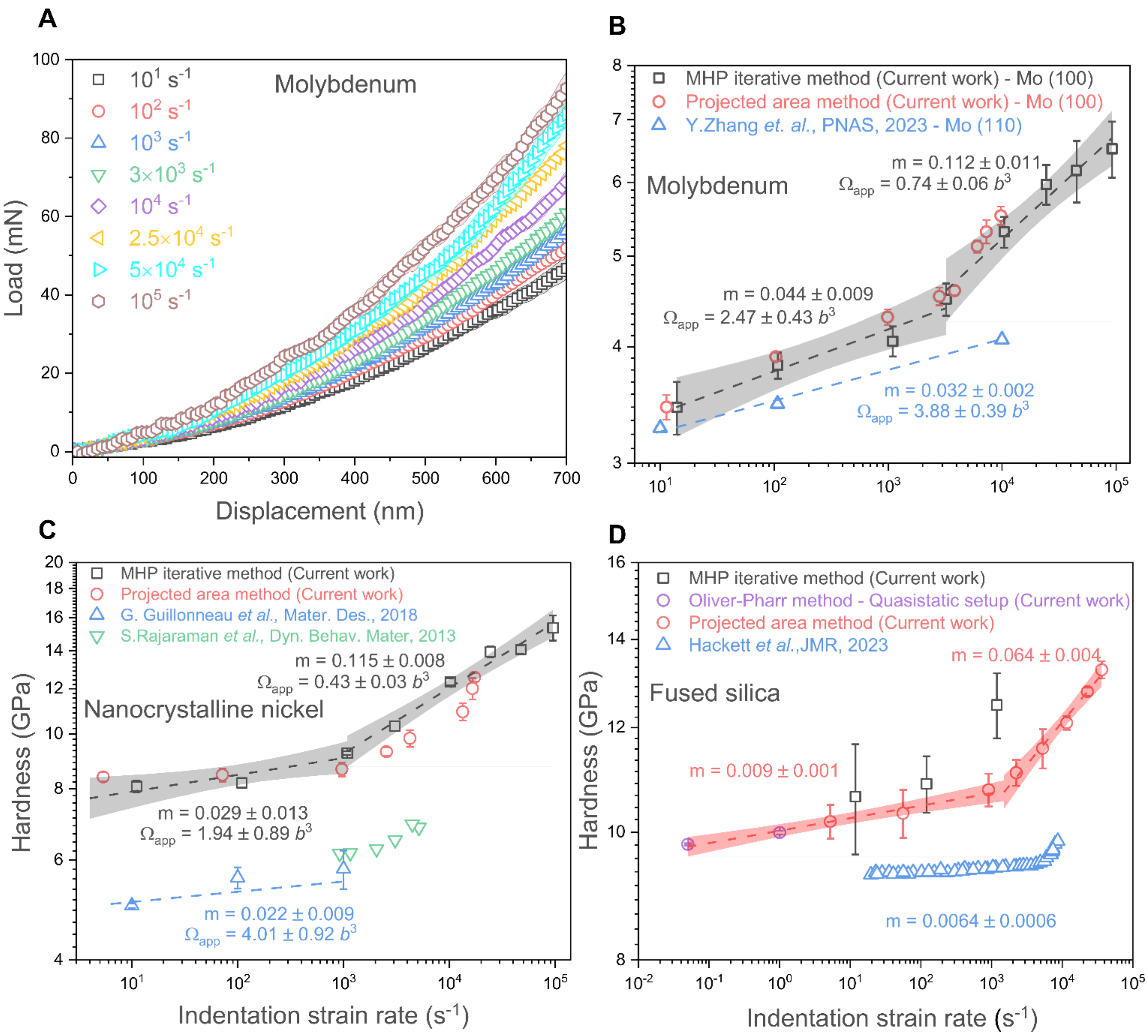

Fig. 2 **Hardness upturn in single crystalline molybdenum, nanocrystalline nickel, and amorphous fused silica**

(**A**) Load-displacement curve for single-crystalline molybdenum after applying all corrections, and hardness values for (**B**) molybdenum, (**C**) nanocrystalline nickel, and (**D**) fused silica, obtained using both the iterative and projected area methods and compared with literature values. Both the axes in (**B-D**) are in log scale. The strain rate sensitivity (m) was calculated using the formula, $\mathrm{m} = \frac{\partial lnH}{\partial ln\dot{\varepsilon}}$, where H is hardness and $\dot{\varepsilon}$ is strain rate. Similarly, the apparent activation volume ($\Omega_{app}$) in figure (**B**) & (**C**) was calculated using the formula, $\Omega_{app} = \sqrt{3}Ck_BT\frac{\partial ln\dot{\varepsilon}}{\partial H}$, where $C$ is the proportionality constant and equal to 3, $k_B$ is the Boltzmann constant, $T$ is temperature, H is hardness and $\dot{\varepsilon}$ is strain rate. The results are expressed in terms of the Burgers vector *b*, which is 0.272 nm for molybdenum and 0.249 nm for nanocrystalline nickel. The bands are with 95% confidence index.

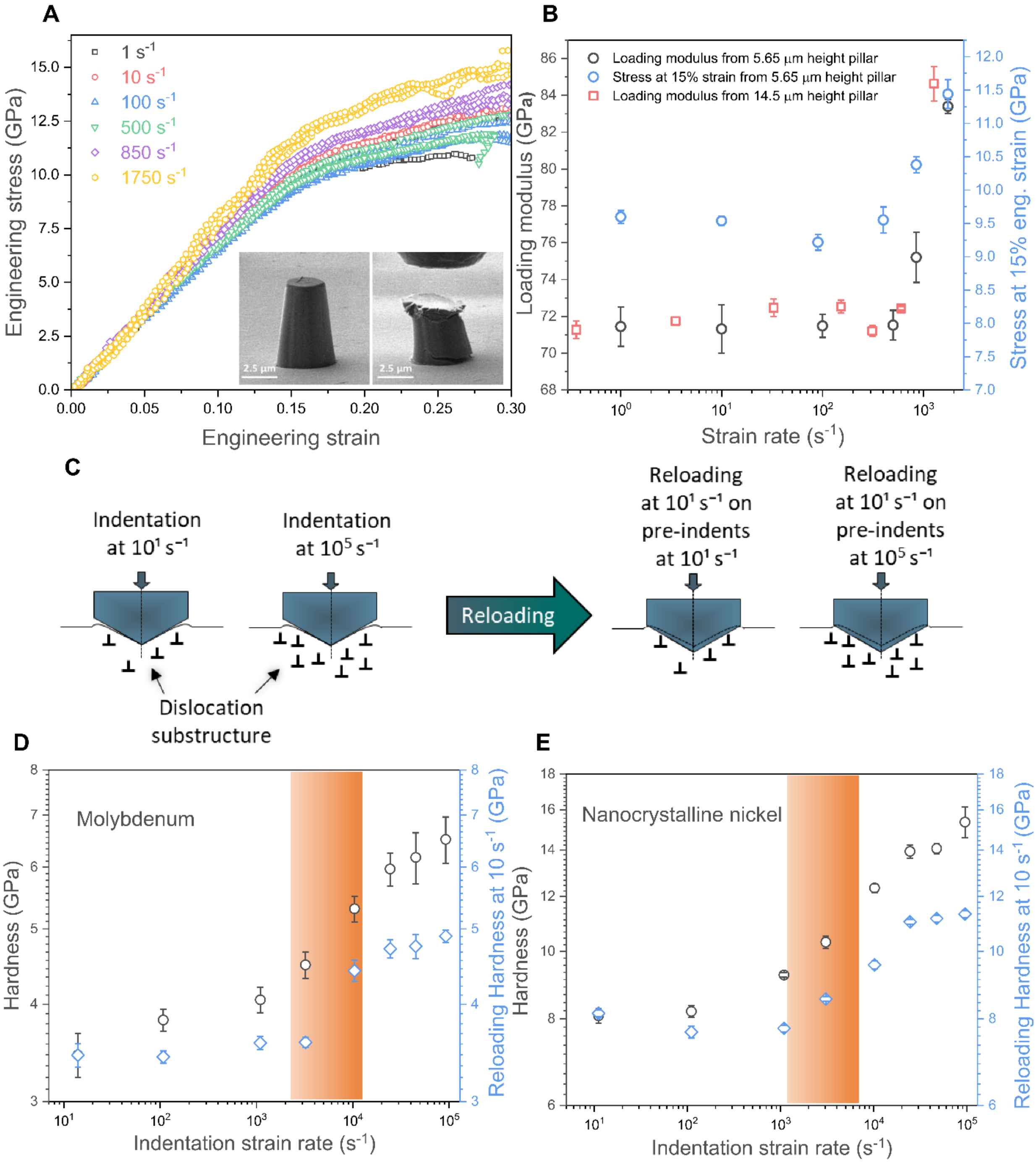

Fig. 3 **Validation of hardness upturn**

**(A)** Shows the engineering stress–strain curves for fused silica micropillars of 5.65 µm height tested at strain rates ranging from 1 $s^{-1}$ to 1750 $s^{-1}$(**B**) shows quantitative analysis of loading modulus (for 5.65 µm and 14.5 µm height fused silica pillar) and flow stress at 15% engineering strain (for 5.65 µm height fused silica pillar), showing significant changes at strain rates of ~$10^3$ $s^{-1}$ and beyond. (**C**) shows the schematic of the reloading process, where all reloads on pre-indents at various strain rates were conducted using a profile of $10^1$ $s^{-1}$. (**D**) & (**E**) shows reloading hardness trends for molybdenum and nanocrystalline nickel, respectively, with hardness vs. strain rate trends overlaid for clarity. The strain rate at which the hardness upturn is observed is marked by an orange band and the hardness values are in log scale.

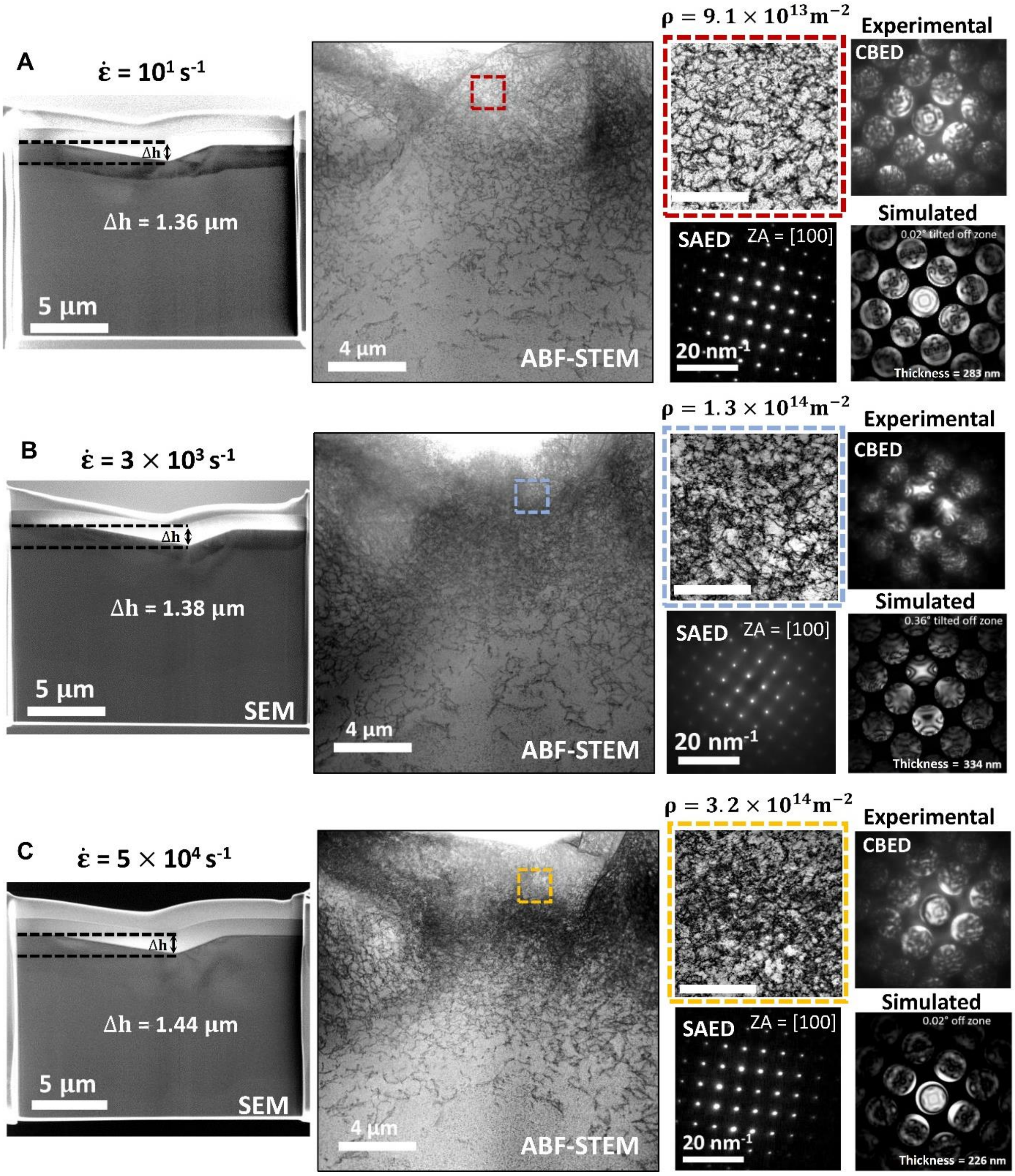

Fig. 4 **Reason for hardness upturn in single crystalline molybdenum**

Scanning electron microscopy (SEM) images, annular bright field scanning transmission electron microscopy (ABF-STEM) images, selected area electron diffraction (SAED), and experimental and simulated convergent beam electron diffraction (CBED) images for molybdenum indents at strain rates of (**A**) $10^1$ s$^{-1}$, (**B**) $3 \times 10^3$ s$^{-1}$, and (**C**) $5 \times 10^4$ s$^{-1}$. The SEM images display indentation depths across strain rates, while the ABF-STEM images highlight the region (approximately 1.5 µm away from the indent tip) used for dislocation density measurements. The scale bar in highlighted regions correspond to 750 nm. The dislocation density values and TEM lamella thickness calculated from the ABF-STEM images and simulated CBED.

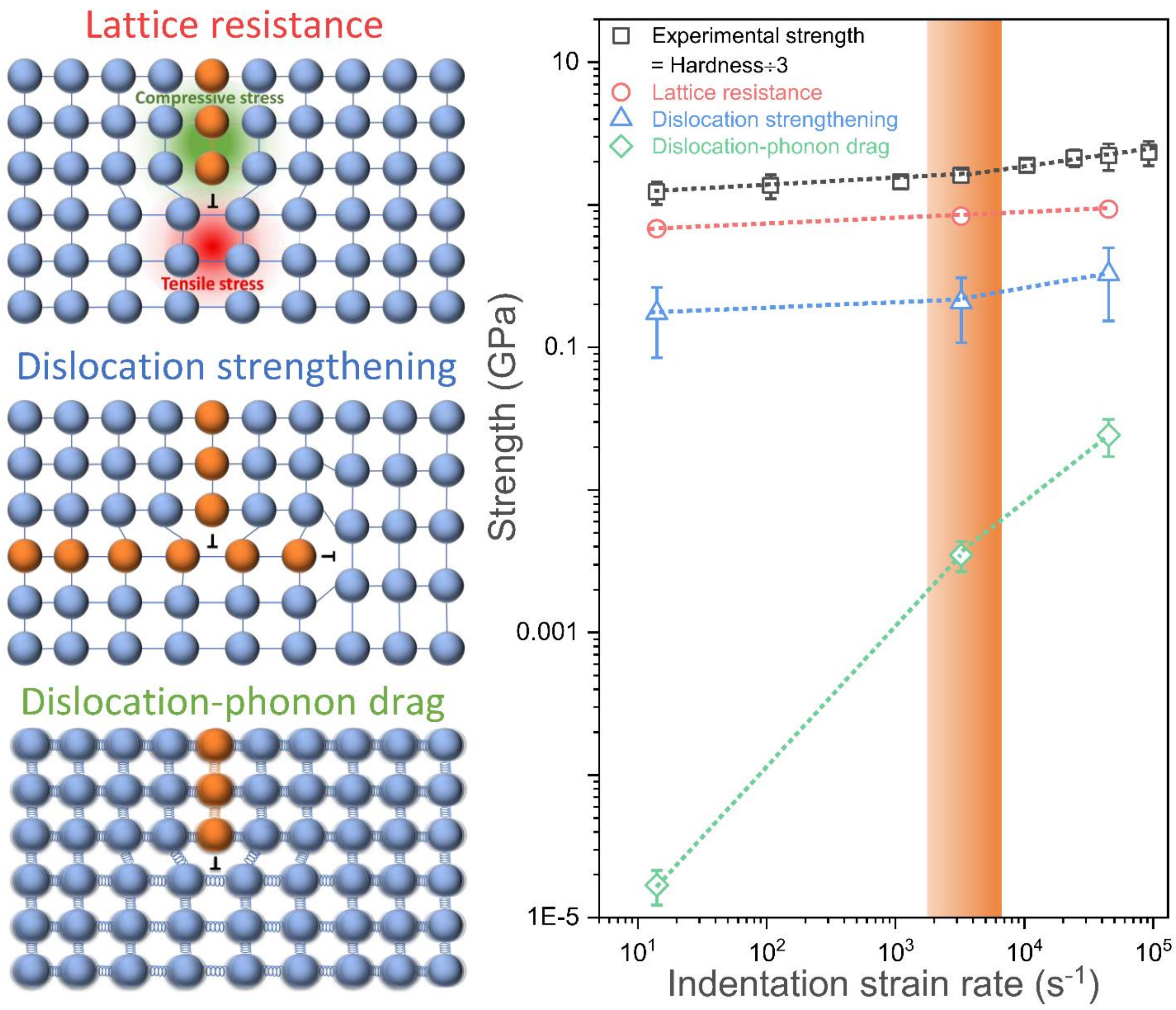

Fig. 5 **Deformation mechanism behind hardness upturn**

The trends of potential strengthening mechanisms for single-crystalline molybdenum, including lattice resistance, dislocation strengthening, and dislocation-phonon drag. The measured hardness values, normalized by dividing by a constraint factor of 2.8, are plotted alongside the calculated strengths for comparison. A guide to the eye is provided in all strengthening mechanisms. The strain rate at which the strength upturn is observed is marked by an orange band and the strength values are in log scale. Schematics illustrating each strengthening mechanism are also included.

# Supplementary information

Filling a gap in materials mechanics: Nanoindentation at high constant strain rates upto $10^5$ $s^{-1}$

Lalith Kumar Bhaskar*[1], Dipali Sonawane[1], Hendrik Holz[1], Jeongin Paeng[1], Peter Schweizer[1], Jing Rao[1], Bárbara Bellón[1], Damian Frey[2], Aloshious Lambai[3,4], Laszlo Pethö[5], Johann Michler[5,6], Jakob Schwiedrzik[7], Gaurav Mohanty[3], Gerhard Dehm*[1], Rajaprakash Ramachandramoorthy*[1]

[1] Max-Planck-Institute for Sustainable Materials, Department of Structure and Micro-/Nano- Mechanics of Materials, Max Planck-Strasse 1, 40237 Düsseldorf, Germany

[2] Alemnis AG, Schorenstrasse 39, 3645 Thun, Switzerland

[3] Materials Science and Environmental Engineering, Faculty of Engineering and Natural Sciences, Tampere University, 33014 Tampere, Finland

[4] Current address: VTT Technical Research Centre of Finland Ltd., Kemistintie 3, FI-02044 Espoo, Finland

[5] Laboratory of Mechanics of Materials and Nanostructures, Empa – Swiss Federal Laboratories for Materials Science and Technology, Feuerwerkerstrasse 39, 3602 Thun, Switzerland

[6] Ecole Polytechnique Fédérale de Lausanne, Institute of Materials (IMX), CH-1015 Lausanne, Switzerland

[7] Laboratory for High Performance Ceramics, Empa – Swiss Federal Laboratories for Materials Science and Technology, Ueberlandstrasse 129, 8600 Dübendorf, Switzerland

*Corresponding author: l.bhaskar@mpier.de; g.dehm@mpie.de; r.ram@mpie.de

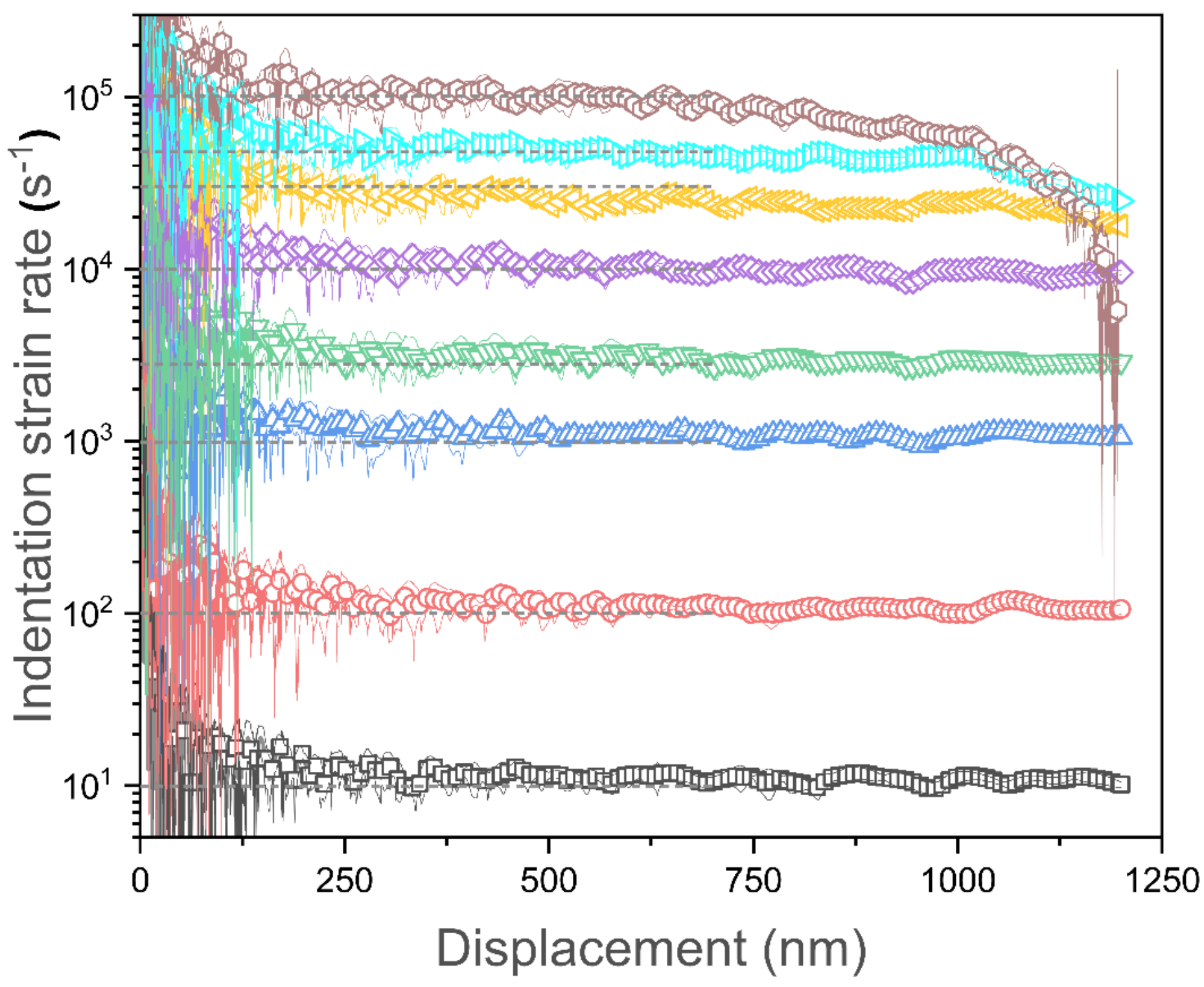

**Fig. S1.**

The strain rate variations as a function of indentation depth.

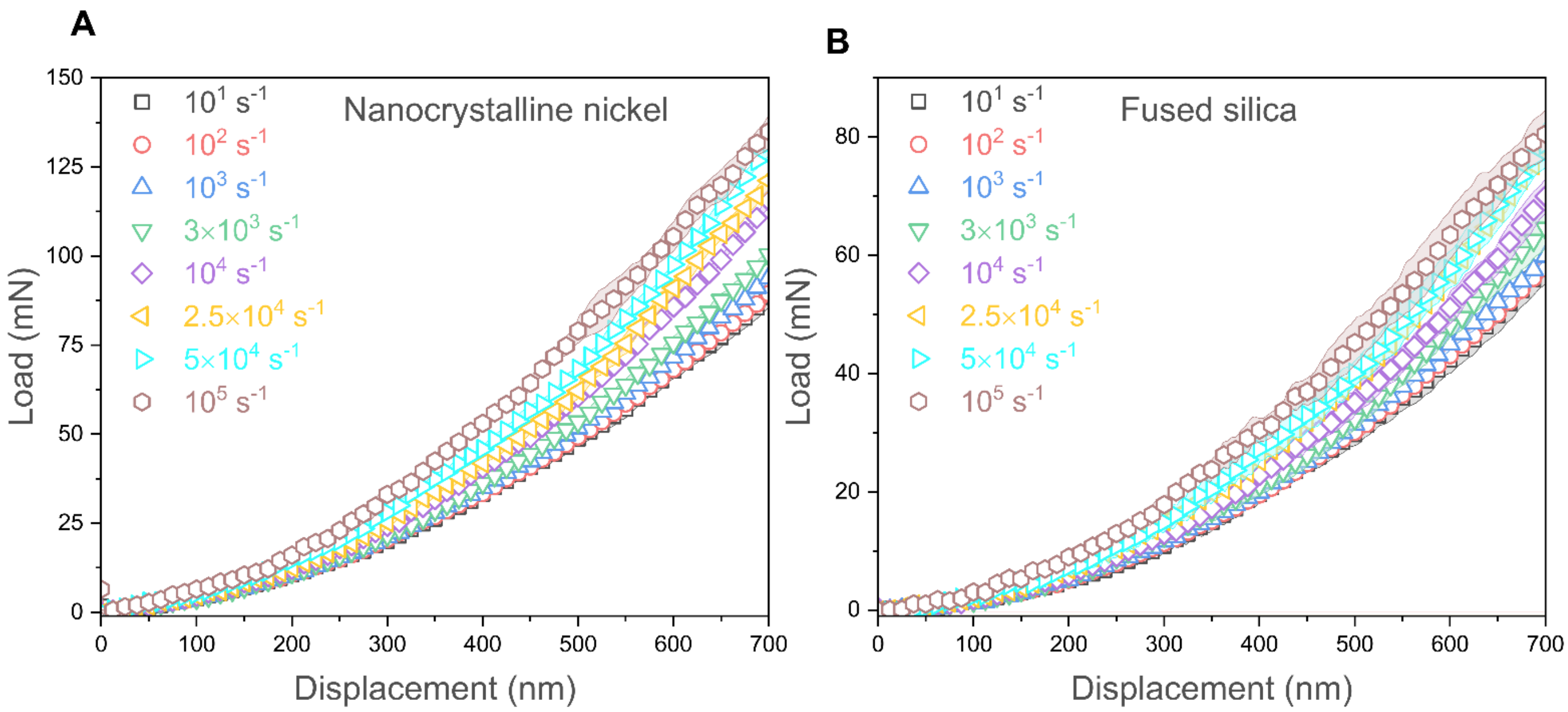

**Fig. S2.**

The load-displacement curve for (**A**) nanocrystalline nickel and (**B**) amorphous fused silica after applying all correction protocols.

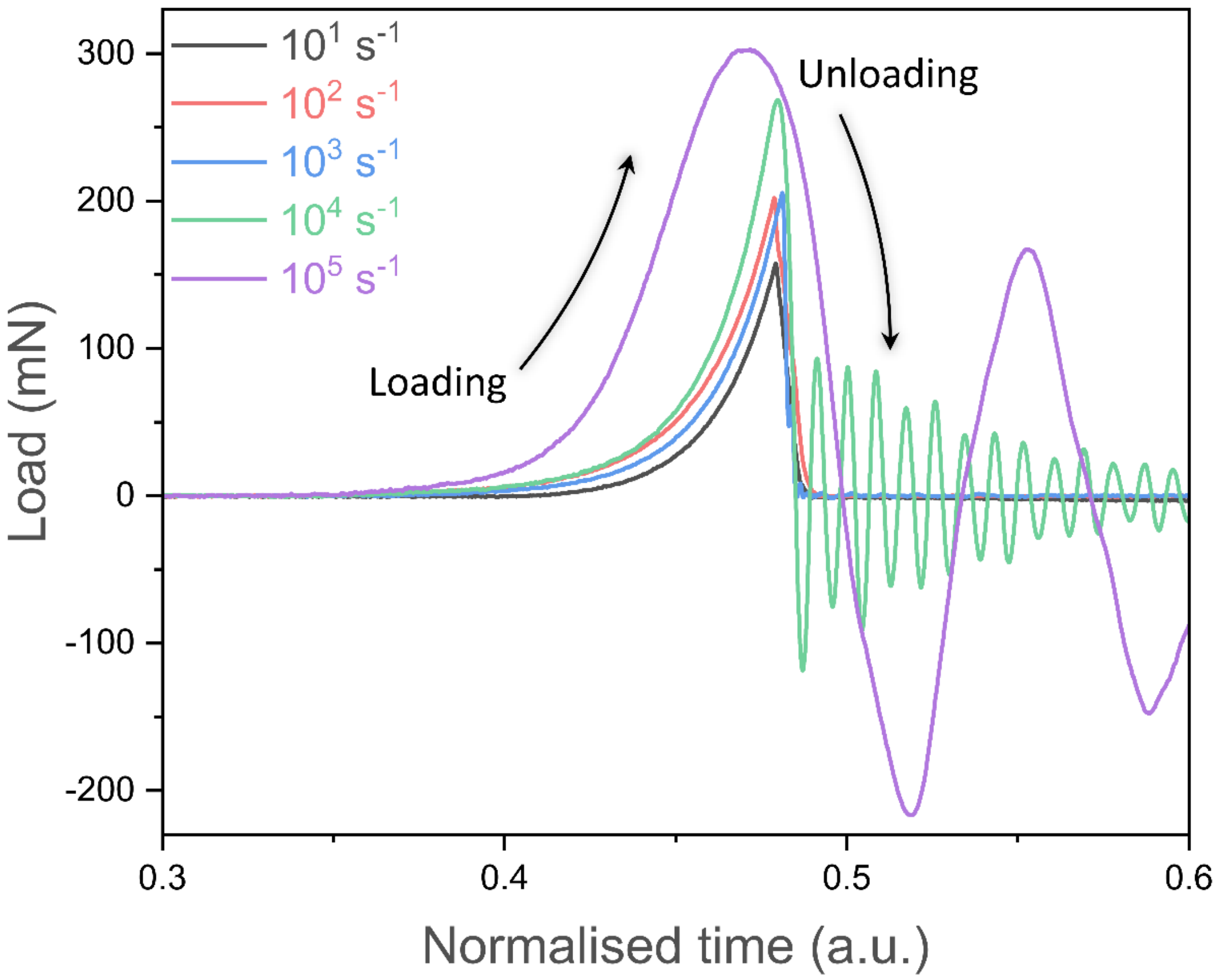

**Fig. S3.**

Representative experimental load-normalised time curves and it can be observed that unloading segments starting from $10^3$ $s^{-1}$ are affected by resonance from various components in the testing platform

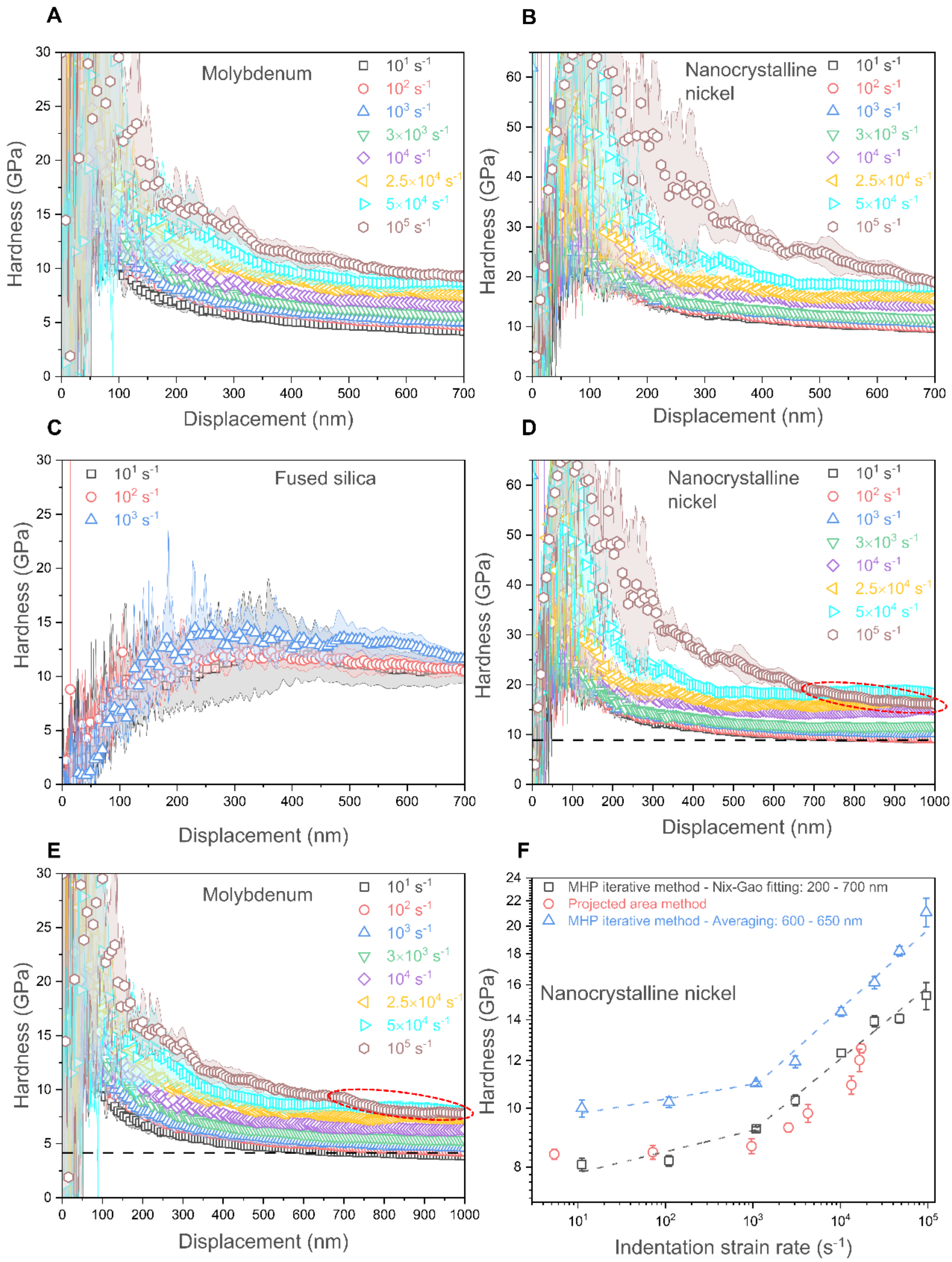

**Fig. S4.**

Hardness vs. displacement plots for **(A)** molybdenum, **(B)** nanocrystalline nickel, and **(C)** fused silica obtained using the MHP iterative method. Each data point includes error bars representing the standard deviation from five independent measurements. Hardness vs. displacement for **(D)** nanocrystalline nickel and **(E)** molybdenum up to a depth of 1000 nm. The decrease in hardness due to decrease in strain rate beyond 700 nm at a strain rate of $10^5$ s$^{-1}$ is highlighted with a dotted red ellipse. **(F)** Comparison of nanocrystalline nickel hardness values determined by three methods: averaging over 600–650 nm, Nix–Gao fitting and extrapolation (200–700 nm), and the projected area method

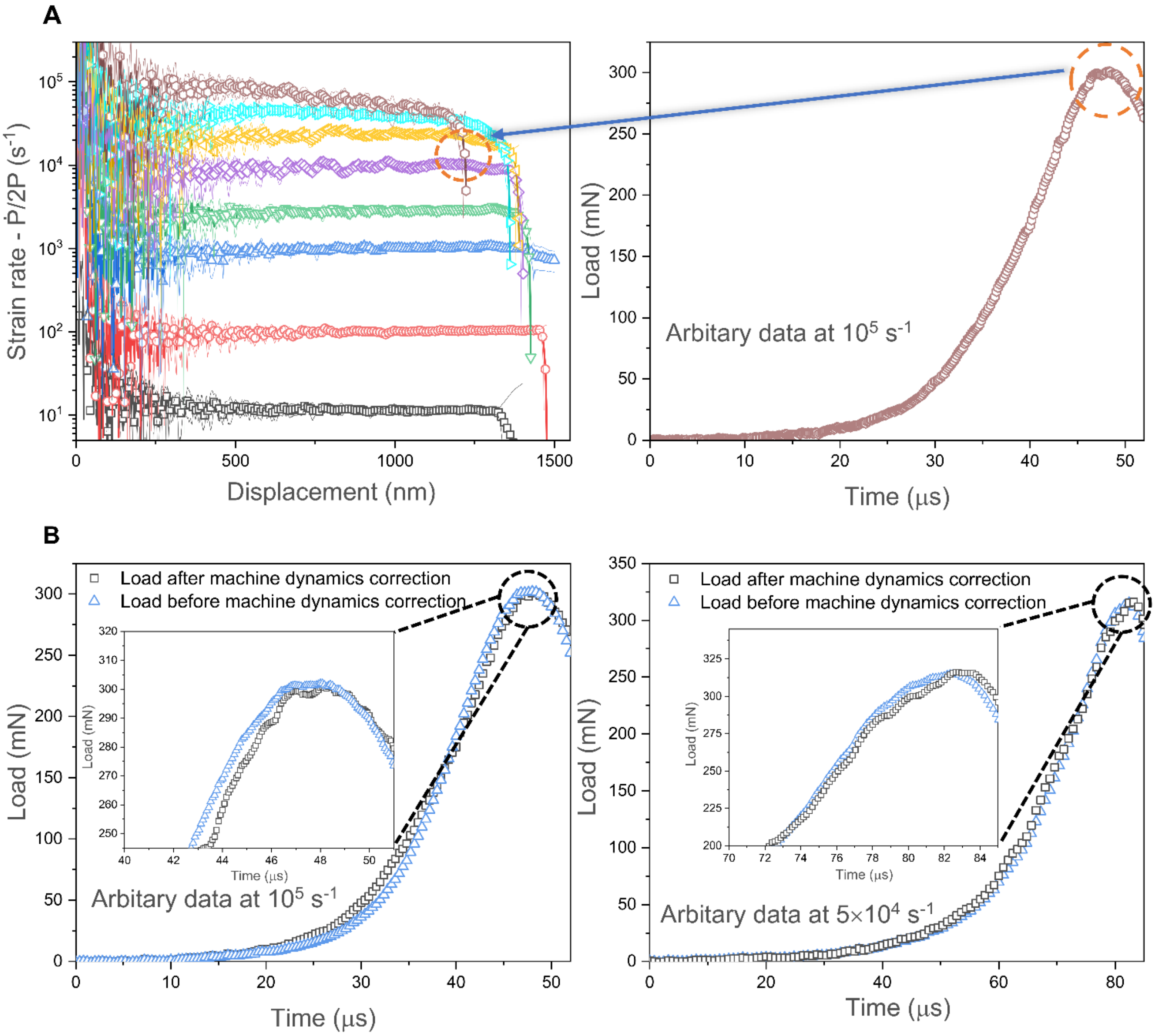

**Fig. S5.**

(**A**) At the highest load, the strain rate falls below the expected/prescribed strain rate, and (**B**) for representative data at $10^5$ s$^{-1}$ and $5\times10^4$ s$^{-1}$, the peak load value remains unchanged after applying all corrections despite changes observed in the loading portion of the curve. Similar trends are observed across all strain rates.

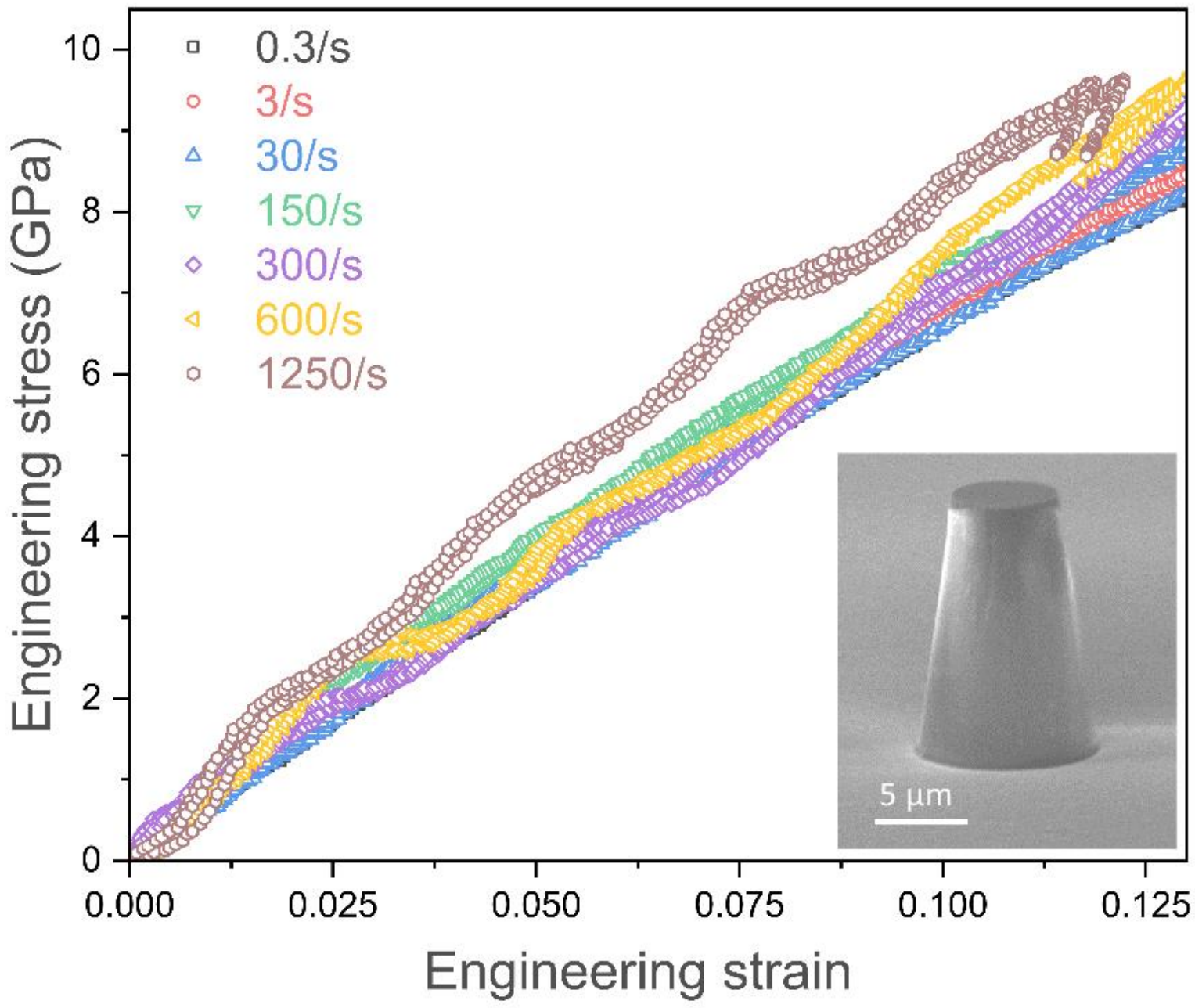

**Fig. S6.**

Shows the engineering stress–strain curves for fused silica micropillars of 14.5 µm height tested at strain rates ranging from 0.3 $s^{-1}$ to 1250 $s^{-1}$. The inset figure shows a representative14.5 µm high fused silica pillar before compression.

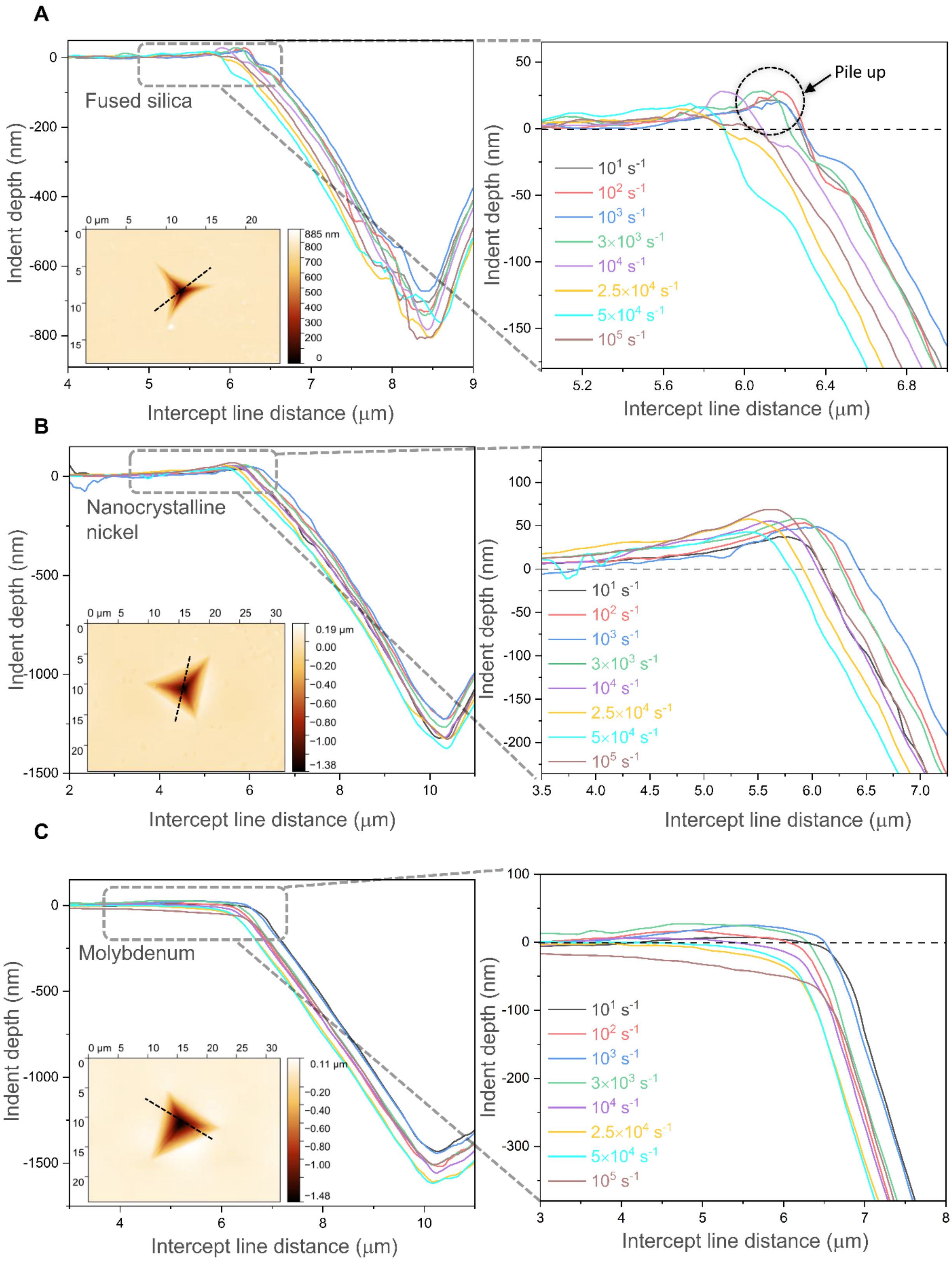

**Fig. S7.**

Cross-sectional line profiles of the indents parallel to one of the indent edges for (**A**) fused silica, (**B**) nanocrystalline nickel and (**C**) molybdenum. The inset figure shows a representative confocal image with the direction along which line profiles were taken.

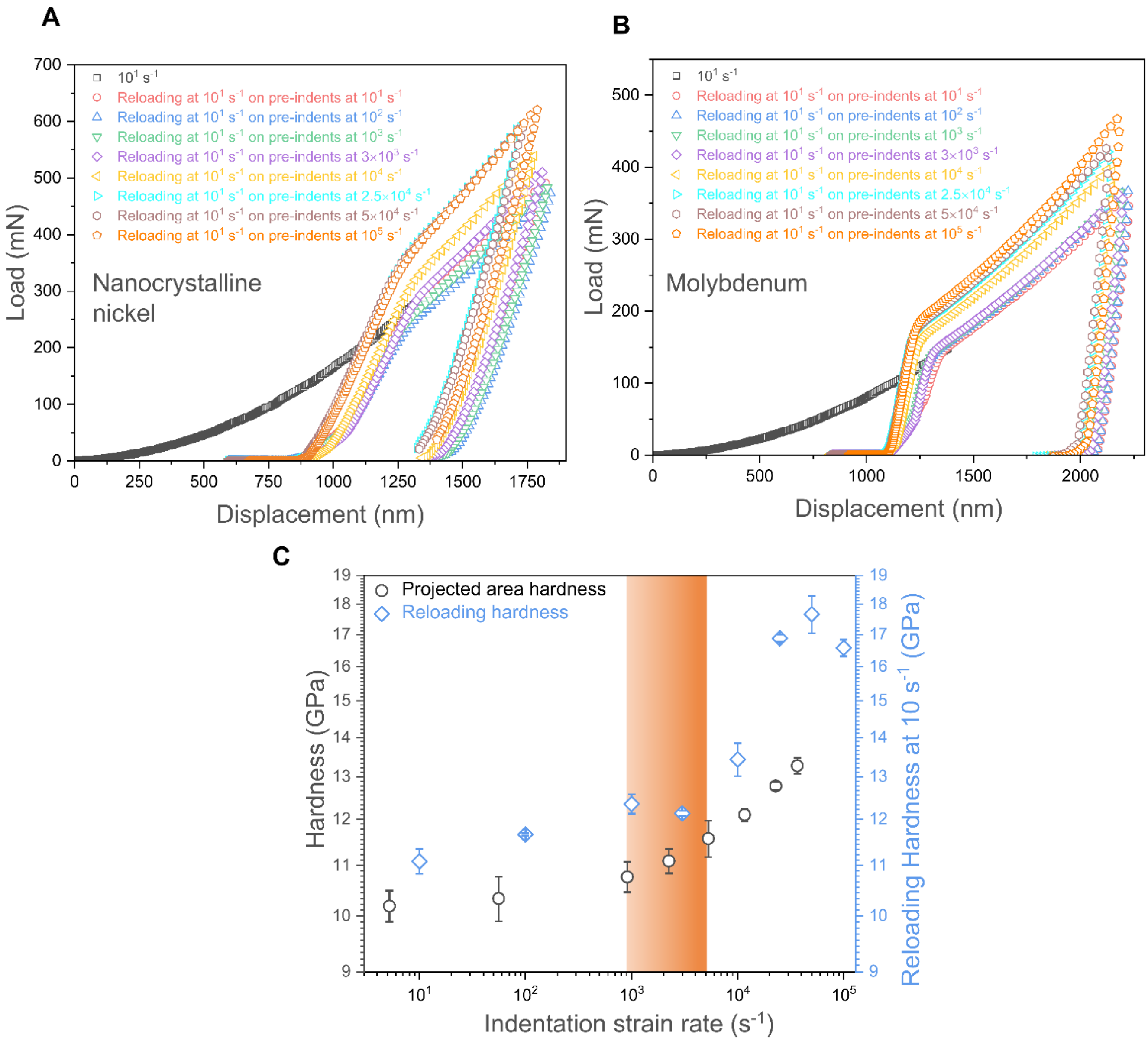

**Fig. S8.**

Reloading load-displacement curves at $10^1$ $s^{-1}$ on pre-existing indents created at various strain rates for (**A**) nanocrystalline nickel and (**B**) molybdenum and (**C**) reloading hardness trends for fused silica, with hardness vs. strain rate trends overlaid for clarity. The strain rate at which the hardness upturn is observed is marked by an orange band and the hardness values are in log scale

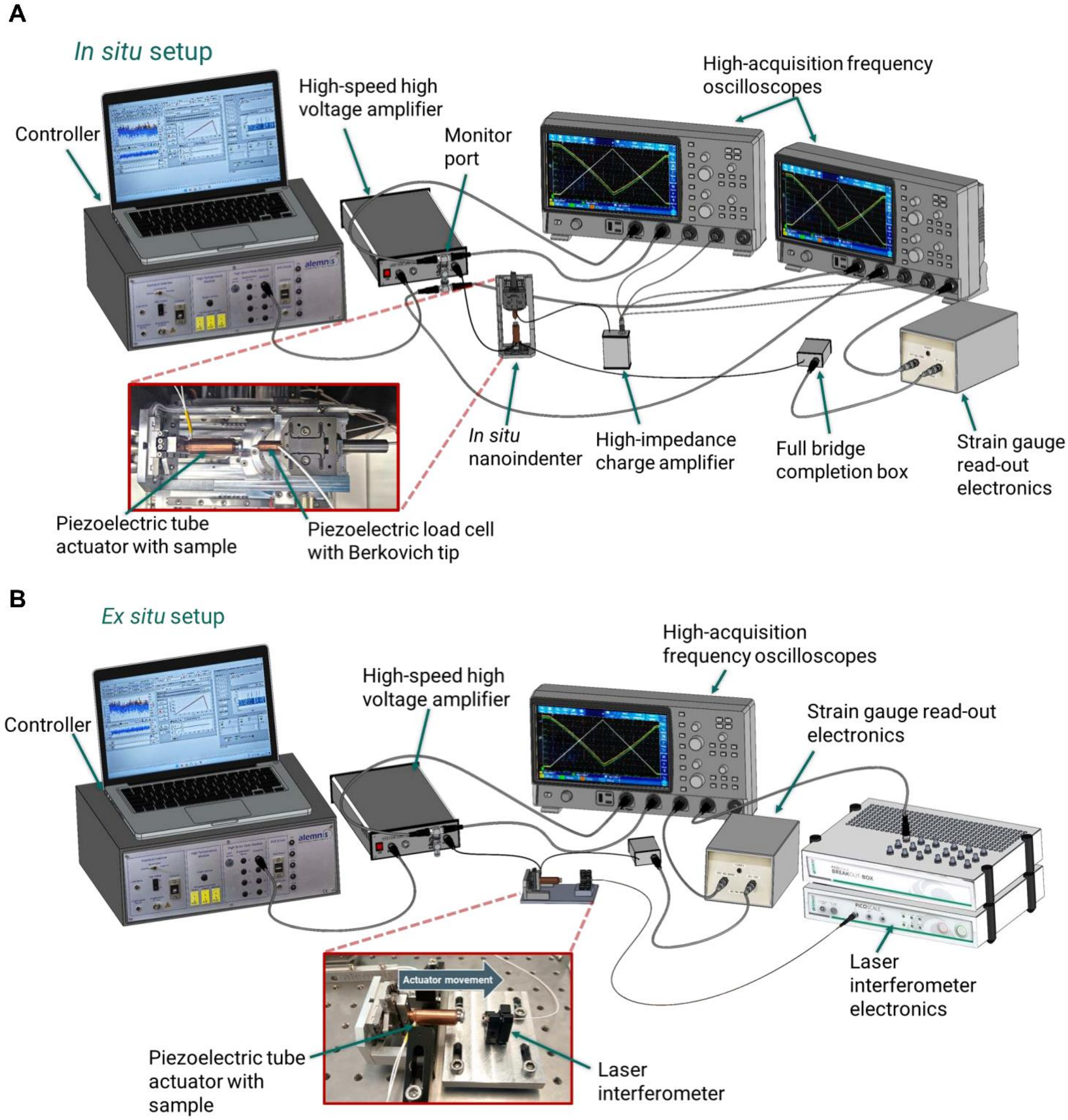

**Fig. S9.**

The (**A**) *in situ* and (**B**) *ex situ* setups with the various components used for performing high constant strain rate indentations.

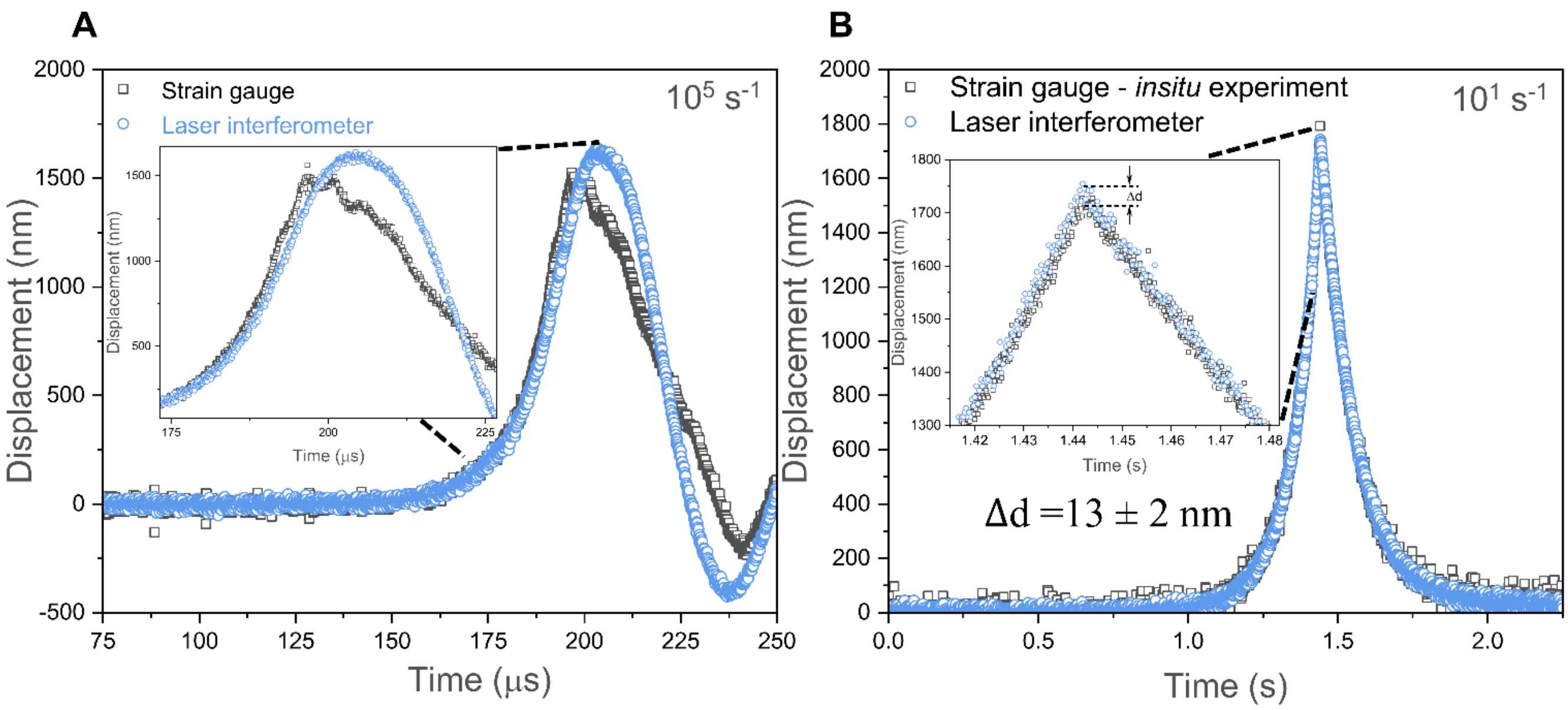

**Fig. S10.**

(**A**) A comparison illustrating that strain gauges mounted on the actuator provide inaccurate displacement readings at very high strain rates ($10^5$ s$^{-1}$) when compared to the laser interferometer; and (**B**) a comparison between *ex situ* displacement measurements from the laser interferometer and *in situ* displacement data recorded during an indentation experiment at $10^1$ s$^{-1}$, showing little to no variation in displacements.

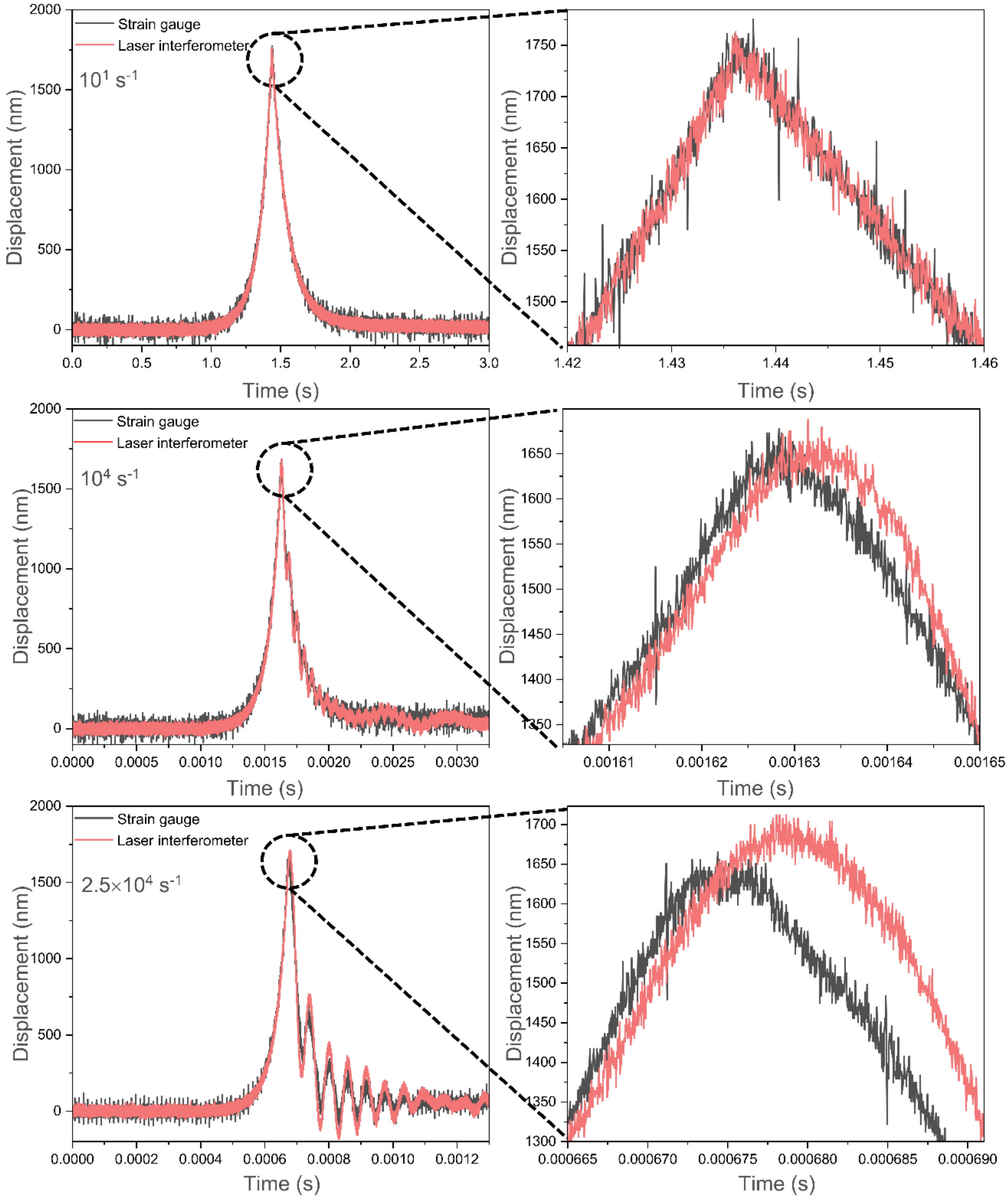

**Fig. S11.**

The displacement data recorded using both the strain gauge and laser interferometer across various strain rates. It is evident that at strain rates exceeding $10^4$ s$^{-1}$, the piezo-resistive strain gauges mounted on the actuator fail to provide accurate data.

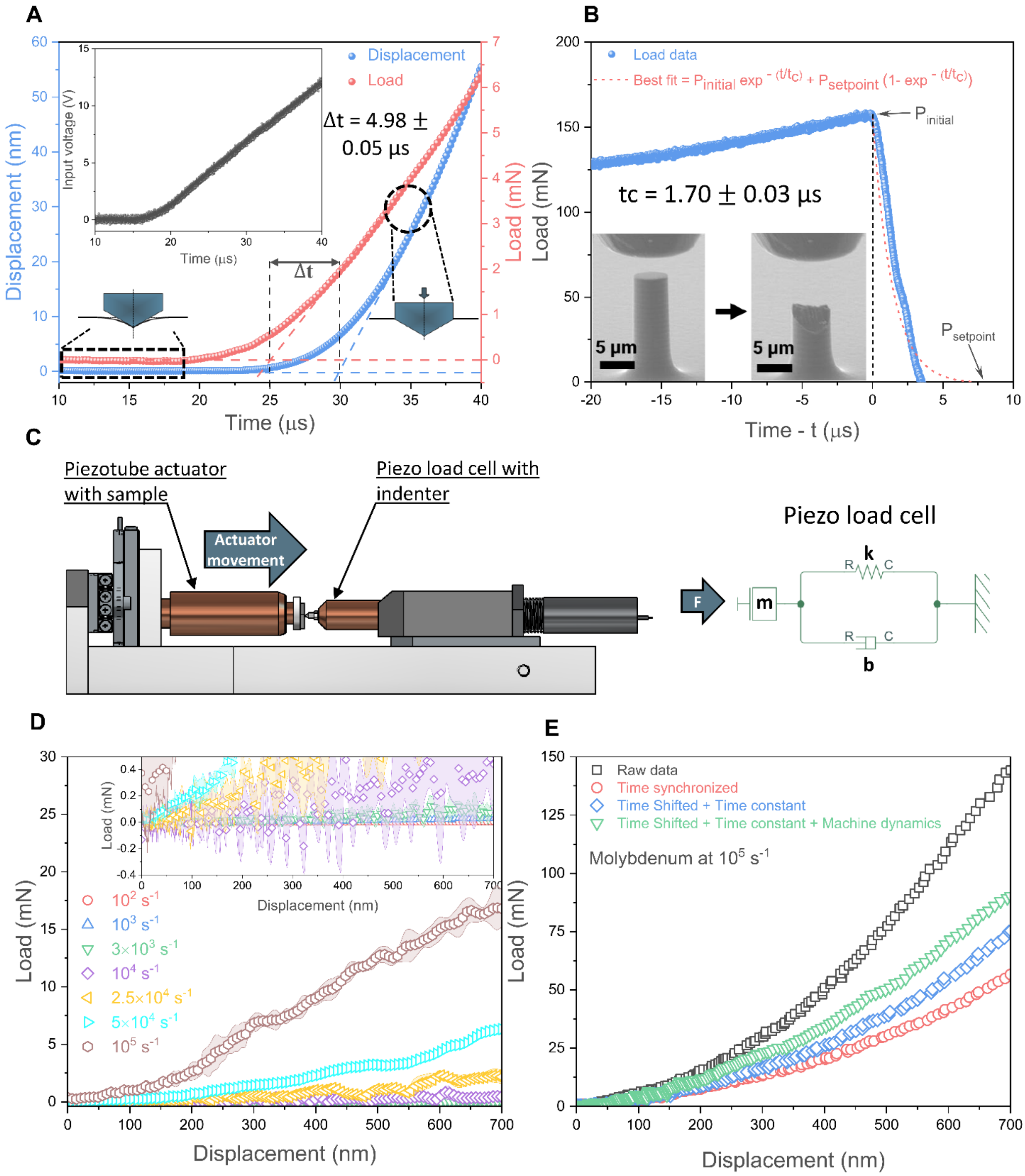

**Fig. S12.**

(**A**) The time lag between load and displacement signals, with inset showing the sharp temporal voltage profile sent to the actuator to capture this lag, and a schematic of the test process. (**B**) Load versus time data for silicon pillar fracture, showing an exponential decay post-fracture rather than an immediate drop (dotted black line); time constant ($t_c$) is calculated using the equation in legend. (**C**) Schematic arrangement of series arrangement of the piezoelectric load cell and piezoelectric tube actuator, with the equivalent Kelvin-Voigt model of the piezoelectric load cell under external force. (**D**) Load contributions from machine dynamics at various strain rates for molybdenum with the inset showing for lower strain rates. (**E**) The cumulative effect of corrections—time synchronization, time constant, and machine dynamics.

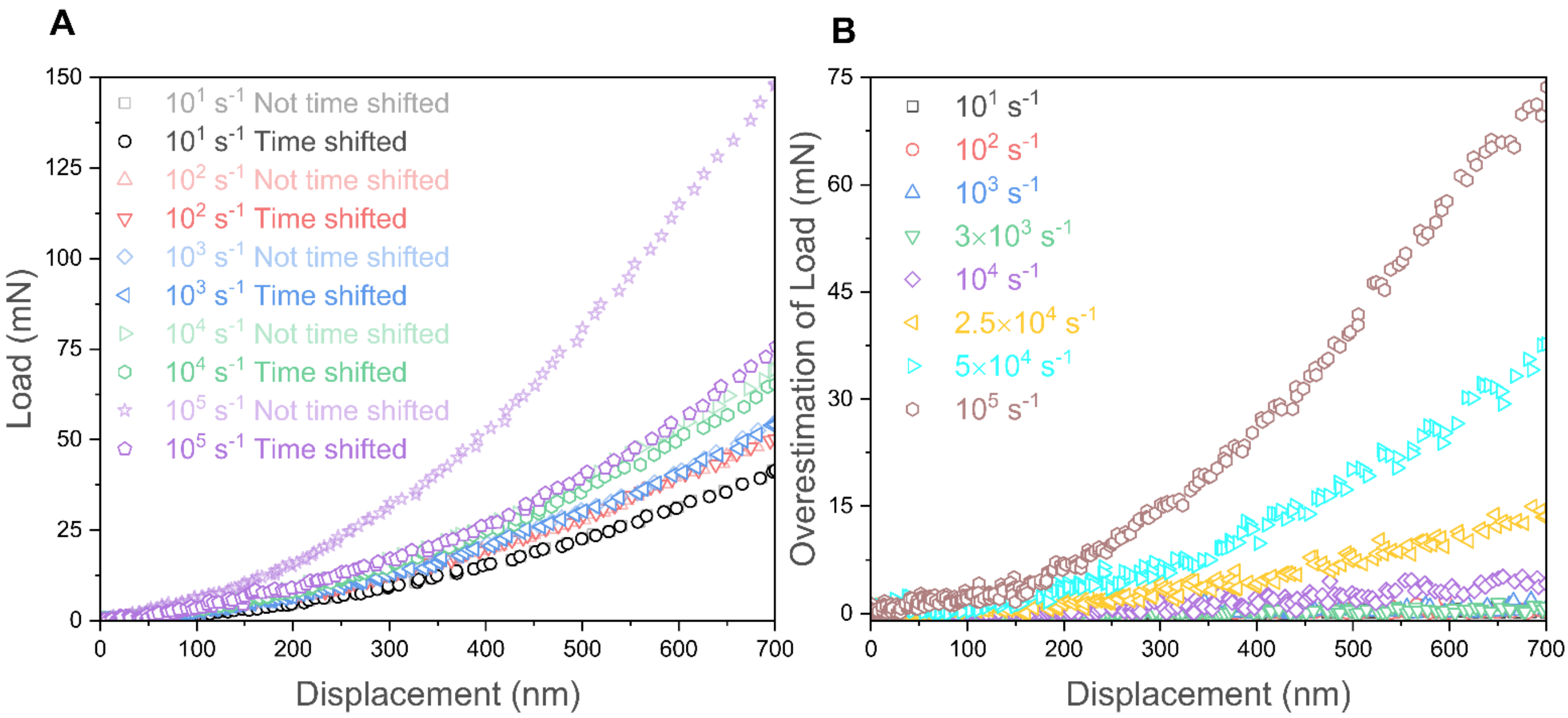

**Fig. S13.**

(**A**) The effect of the 4.98 ± 0.05 μs time lag on the load-displacement data for molybdenum at different strain rates, while (**B**) highlights the load overestimation as a function of displacement across all tested strain rates if time synchronization is not applied. At strain rates exceeding $10^3$ s⁻¹, the error becomes noticeable and starts to propagate due to the time lag.

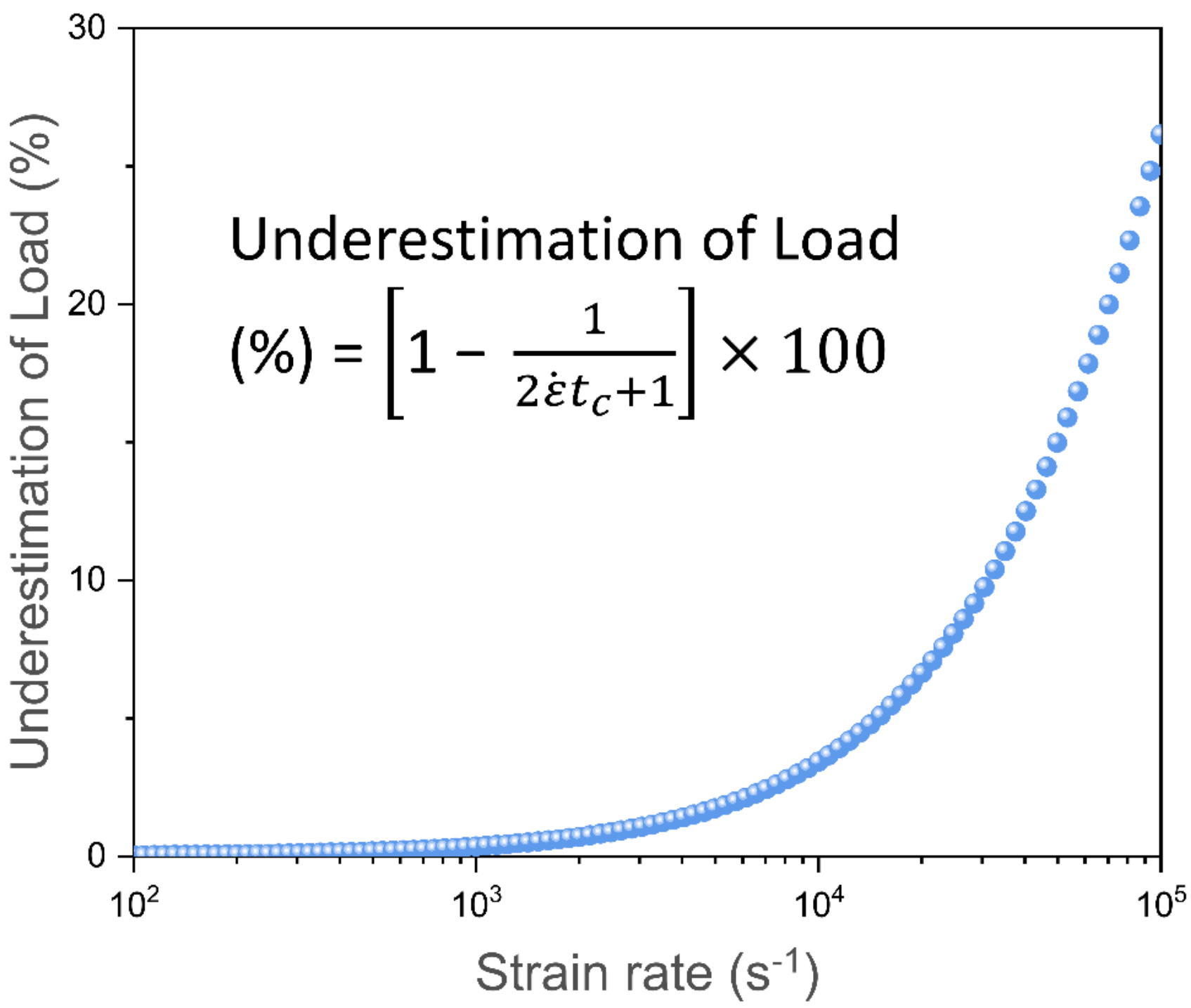

**Fig. S14.**

The percentage underestimation of the load signal as a function of strain rate when the time constant ($t_c$) is not considered.

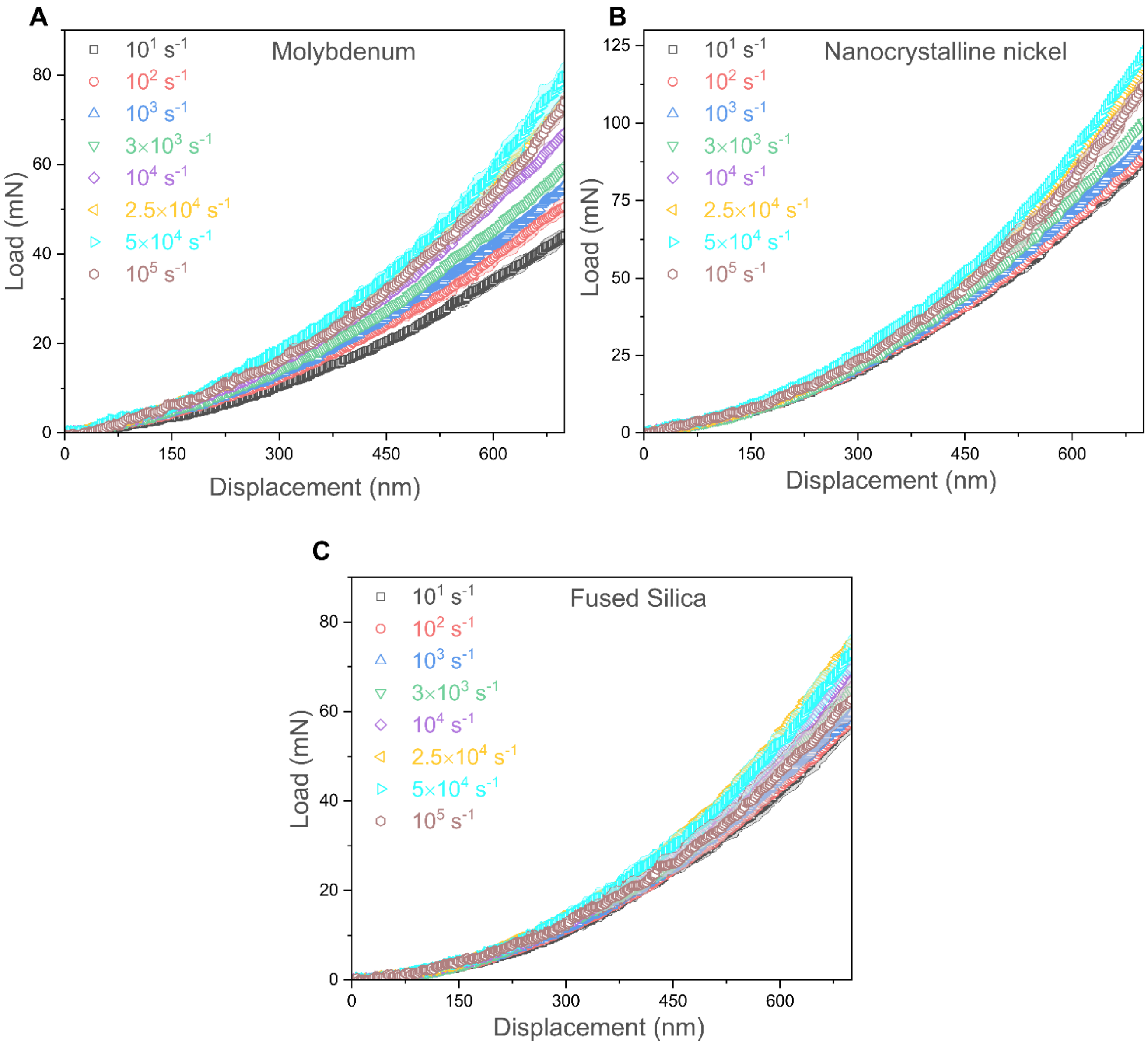

**Fig. S15.**

The load-displacement curves after applying time synchronization, time constant correction, compliance correction, and zero-displacement correction for (**A**) molybdenum, (**B**) nanocrystalline nickel, and (**C**) fused silica.

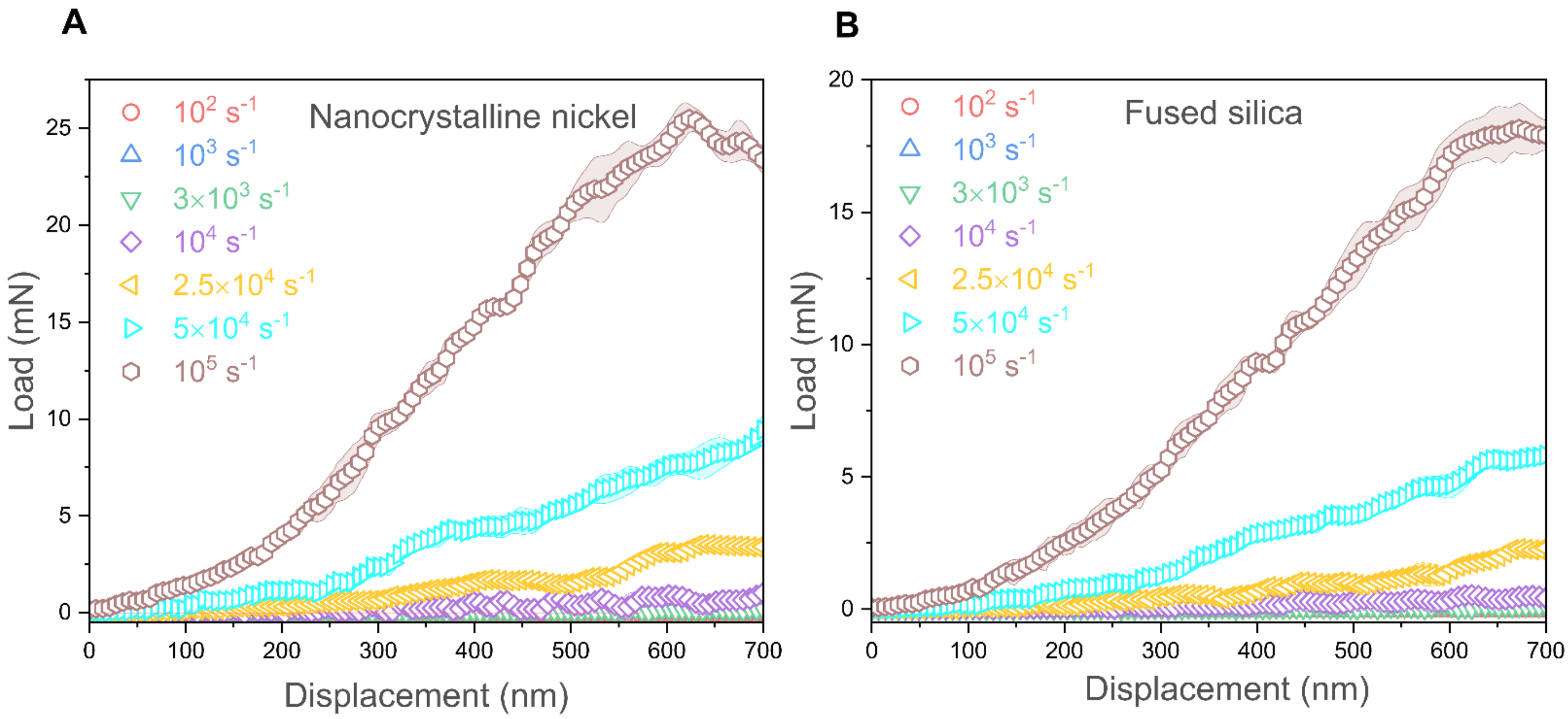

**Fig. S16.**

The load contributions from machine dynamics at various strain rates for (**A**) nanocrystalline nickel and (**B**) fused silica.

### *S1 Nix-Gao fitting for nanocrystalline nickel*

Although indentation size effects are not expected for nanocrystalline nickel[1], Nix–Gao fitting was still performed, because hardness did not saturate at depths below 700 nm—a behavior similar to what has been reported in prior studies[2]. Only beyond 700 nm did hardness show saturation. Supporting evidence can be seen in fig. S4D and S4E, which display hardness data for both nanocrystalline nickel and molybdenum up to 1000 nm. A guide-for-the-eye horizontal dotted line helps visualize that hardness for nanocrystalline nickel stabilizes beyond 700 nm, while molybdenum does not, due to indentation size effect. Additionally, fig. S4F presents hardness values for nanocrystalline nickel obtained both by averaging over depths of 600–650 nm and by applying the Nix–Gao model (200–700 nm range) with extrapolation. While the relative trends match, the absolute values from the averaging method are higher. Using greater depths for averaging is not feasible at the highest strain rate ($10^5$ s$^{-1}$), as strain rate drops at depths greater than 700 nm (fig. S1), which, as a result, is reflected in hardness (see fig. S4D & S4E).

### ***S2 Strengthening mechanism***

*Lattice resistance strengthening*

In most polycrystalline materials, the motion of a dislocation is constrained by its interaction with the atomic structure known as the Peierls force or lattice resistance, this arises from fluctuations in the dislocation's energy as it moves through the crystal lattice. The magnitude and periodicity of these fluctuations are influenced by the strength and spacing of the interatomic or intermolecular bonds. It is given by equation,

$$\sigma_l = \left[1 - \left(\frac{kT}{g_0 \mu_0 b^3} \ln\left(\frac{\dot{\varepsilon}_0}{\dot{\varepsilon}}\right)\right)^{1/q}\right]^{1/p} \sigma_0 \qquad (1)$$

Here, $g_0$ is a constant related to the strength of the obstacles, determined by the microstructure. In this case, it represents the lattice resistance with a value of 0.1. $b$ is the magnitude of the Burgers vector, equal to 0.272 nm, and $\mu_0$ is the shear modulus with a value of 134 GPa. The constants $p$ and $q$ are exponents that describe the shape and spacing of energy barriers associated with the obstacles, with values of 3/4 and 4/3, respectively. $k$ represents Boltzmann’s constant, $T$ is the temperature (300 K in this case), $\sigma_0$ is the stress required to overcome the relevant short-range barrier at 0 K, valued at 0.8844 GPa, and $\dot{\varepsilon}_0$ is the reference strain rate, set at $10^{11}$ $s^{-1}$. All constant values are sourced from Harold and Ashby, *Deformation of Mechanism Map* [3].

*Dislocation strengthening*

In addition to lattice resistance, long-range obstacles like far-field forest dislocations in a single crystalline material can create a potential field that hinders dislocation motion. The interaction strengthening is given by the following Taylor equation,

$$\sigma_{disl} = \alpha_{disl} \mu_0 b \sqrt{\rho_t} \qquad (2)$$

Here, $\alpha_{disl}$ is the dislocation-dislocation interaction parameter and is equal to 0.5 and $\rho_t$ is the total dislocation density in the sample and is obtained from the TEM analysis.

*Dislocation-Phonon drag strengthening*

When the applied stress is sufficiently high, dislocations can bypass obstacles continuously, accelerating to high steady-state velocities, where dislocation interacts with, phonons and electrons and which give rise to strengthening. The increase in strength due to phonon drag is given by the equation,

$$\sigma_d = \frac{B_d}{\rho_m b^2}\dot{\varepsilon} \qquad (3)$$

$$B_d = \frac{3kT}{5 \cdot Ca^3}(1.8b) \qquad (4)$$

Here, $\rho_m$ represents the mobile dislocation density, which is typically a fraction of the total dislocation density[4] . For molybdenum, a BCC material with higher lattice resistance, $\rho_m$ is assumed to be, $\rho_m = 10^{-3} \times \rho_t$. Additionally, $C$ denotes the shear wave velocity, which for molybdenum is 3500 m/s. Further details on this formulation can be found in I. Dowding *et al.* [5].

***S3 Machine dynamics – Methodology and error analysis***

As the strain rate increases, it is clear that the force applied on the piezoelectric load cell is fast enough to encounter inertial resistance. Due to this inertial resistance, the compression of the load cell is smaller than expected, meaning the charges output and consequently the measured load ($P_{measured}$) is underestimated compared to the actual true load ($P_{true}$). Hence, to obtain the true load, the measured load must be compensated appropriately with the loads that were lost due to the inertial resistance, as given in Equation 5 below:

$$P_{true} = P_{measured} - (-m\ddot{x} - b\dot{x}) \qquad (5)$$

Here, $\ddot{x}$ and $\ddot{x}$ represents the acceleration and velocity of the piezoelectric load cell respectively. In order to calculate the acceleration and velocity one needs to determine the compressive displacement ($x$) of the piezoelectric load cell due to external force (F) acting on the load cell. From classical mechanics, internal second body compressive displacement ($x$) of the piezoelectric load cell can be related to external force through the following relation,

$$F = kx \qquad (6)$$

In the literature [6], for systems with similar configurations, a common simplification to evaluate the compressive displacement ($x$) has been to assume that the measured load ($P_{measured}$) is same as the external force (F). However, it is clear that the inertial resistance of the load cells does influence the measured load, which would affect the calculated internal compressive displacement ($x$) of the load cell. Therefore, in this study, a different approach was implemented by using a *mock load* ($F_{mock}$) rather than the measured load ($P_{measured}$) to obtain the compressive displacement of the load cell and consequently the required compensation to obtain the true load ($P_{true}$) accurately.

However, before determining the internal compressive displacement ($x$) of the second body (piezoelectric load cell), several other variables must be identified, including the damping coefficient ($b$), elastic spring stiffness ($k$), and mass ($m$). To determine the total mass, the masses of the piezo element in the load cell, the diamond Berkovich tip, and the connector interface between the indenter and the piezo element must be identified. Each component was individually weighed in a precision balance: the mass of the diamond Berkovich tip = 0.1 g, the connector interface = 0.05 g and the piezo element of the load cell = 0.265 g. However, from classical mechanics, when a mass undergoes oscillation or rotation about one end, its contribution to the system's inertia can be approximated by concentrating a fraction of the total mass at a specific location—typically one-third of the mass in cases resembling a cantilevered beam [7]. This simplification captures the dynamic effect of the distributed mass by representing it as a point mass located at an equivalent position, determined by the system's geometry and mode of vibration. In this study, only the piezo element of the load cell is actively oscillating. Therefore, its equivalent mass is considered in the dynamic analysis. This is justified by the fact that, for the fundamental bending mode of a uniform cantilever beam, the effective mass at the free end is well-established to be one-third of the total mass and is given by, $m_{eff} = \frac{1}{3}m.$ This approximation effectively accounts for the portion of the piezo's mass that contributes to dynamic motion.

Therefore, the total mass ($m$) is calculated as follows:

$$Total\ mass\ (m) = \frac{1}{3}\ mass\ of\ piezo\ element + mass\ of\ connector\ interface + mass\ of\ diamond\ Berkovich\ tip \qquad (7)$$

$$\therefore m = \frac{0.265}{3} + 0.05 + 0.1 = 0.238\ \text{g}$$

The stiffness ($k$) of a cylinder can be estimated by the following equation,

$$Stiffness\ (k) = \frac{EA}{L} \qquad (8)$$

Here, the elastic modulus ($E$) of the piezoelectric cylinder used in the load cell was obtained from the supplier (PI Ceramic GmbH). The stiffness matrix of the piezo as given by the supplier is,

$C_{11}$ = 133.2 GPa

$C_{12}$ = 87.15 GPa

$C_{13}$ = 86.19 GPa

$C_{33}$ = 120 GPa

$C_{44}$ = 21.76 GPa

$C_{55}$ = 21.76 GPa

$C_{66}$ = 23.02 GPa

Then using the ELATE tool [8] the elastic modulus ($E$) was calculated to be 60.753 GPa.

The area ($A$) of the piezoelectric tube is given by,

$$Area\ (A) = \frac{\pi}{4}(Outer\ diameter^2 - Inner\ diameter^2)$$

$$= \frac{\pi}{4}((3.2E-3)^2 - (2.2E-3)^2) = 4.24E-6\ m^2 \qquad (9)$$

And the length of the tube ($L$) was measured to be 8mm. Substituting all the values in equation (6) the stiffness ($k$) was calculated to be 3.22E7 N/m.

The damping coefficient ($b$) of the system was determined by calculating the damping ratio ($\xi$) of the piezoelectric load cell. For this, the free body oscillation of the load cell after a quick excitation was collected. Fig. S17 shows the free body oscillation of the piezoelectric load cell after an indentation experiment which was carried out at $5\times10^4$ $s^{-1}$. The damping ratio ($\xi$) is given by the following relation,

$$\xi = \frac{b}{C_c} \qquad (10)$$

Where $C_c$ is the critical damping coefficient and is given by,

$$C_c = 2\sqrt{k \times m} \qquad (11)$$

Substituting for stiffness ($k$) and mass ($m$) the critical damping coefficient ($C_c$) was calculated to be 175.2 $\frac{Ns}{m}$. The damping ratio ($\xi$) in turn is related to the logarithmic decrement ($\delta$) by the following relation [7],

$$\xi = \frac{1}{\sqrt{(1 + \left(\frac{2\pi}{\delta}\right)^2}} \qquad (12)$$

And logarithmic decrement ($\delta$) in turn is given by,

$$\delta = \ln\frac{x_1}{x_2} \qquad (13)$$

where $x_1$ and $x_2$ are amplitudes of two successive oscillations as shown in fig. S17.

Using equations (10-13) the damping coefficient of the piezoelectric load cell was calculated to be 10.2 $\frac{Ns}{m}$.

In this work, a novel approach was employed to determine accurately the external force ($F$) on the load cell. In this work, in total eight strain rates were tested - $10^1$, $10^2$, $10^3$, $3\times10^3$, $10^4$, $2.5\times10^4$, $5\times10^4$, and $10^5$ $s^{-1}$. For the indentation experiment at $10^1$ $s^{-1}$, which has a duration of about 0.5 seconds, the influence due to inertial resistance from the load cell is minimal, so no inertial correction was needed. For strain rates higher than $10^2$ $s^{-1}$ a correction owing to the inertial resistance of the load cell was applied. The following procedure was followed to compensate for the inertial resistance of the load cell based underestimation of the true load at strain rates beyond $10^2$ $s^{-1}$.

A machine dynamics model that represents the sample as a simple non-linear spring and piezoelectric load cell as spring, mass, and dashpot was built. Given that the load-displacement response of the sample at the high strain rates is significantly affected by the inertial resistance of the load cell, the spring stiffness of the sample is obtained by fitting the load-displacement response of the sample at a preceding lower strain rate that is already corrected for machine dynamics. The sample stiffness is multiplied by the actuation displacement, corresponding to the current strain rate, to obtain the *mock load*. The *mock load* is then used to obtain the compressive displacement ($x$) of the load cell using Equation (6). Subsequently, by differentiation with respect to time, the velocity and acceleration of the load cell were also obtained. Using these values in Equation (5), the true load can be calculated. The iterative case study following this protocol starting from 10 $s^{-1}$ till $10^3$ $s^{-1}$ is given below as an example:

The load ($P$)- displacement ($h$) curve of the $10^1$ $s^{-1}$ (slow enough to be not affected by inertial resistance of the load cell) experiment was fit using a quadratic function, $P = C_0h^2 + C_1h$. By using the parameters $C_0$ and $C_1$ along with the compliance-corrected displacement from the $10^2$ $s^{-1}$ experiment, a *mock load* was calculated. This *mock load* ($F_{mock}$) was then used to calculate the internal compressive displacement ($x$) of the piezoelectric load cell using Equation (6) and by differentiating ($x$), the acceleration $\ddot{x}$ and velocity $\dot{x}$ were obtained. By substituting these values into Equation (5), the true load response ($P_{true}$) at $10^2$ $s^{-1}$ corrected for inertial resistance of the load cell could be obtained.

Similarly, for correcting the machine dynamics for experiments carried out at $10^3$ $s^{-1}$, the machine dynamics-corrected load-displacement curve of the $10^2$ $s^{-1}$ experiment was fitted using the quadratic function to obtain updated $C_0$ and $C_1$ parameters. These parameters, along with the compliance-corrected displacement from the $10^3$ $s^{-1}$ experiment, were used to generate a *mock load* that provided an accurate measure of the load cell's compressive displacement ($x$). Using this, the underestimation of load due to the inertial resistance of the load was calculated, and the load-displacement response at $10^3$ $s^{-1}$ was corrected for machine dynamics. This same procedure was repeated for each higher strain rate.

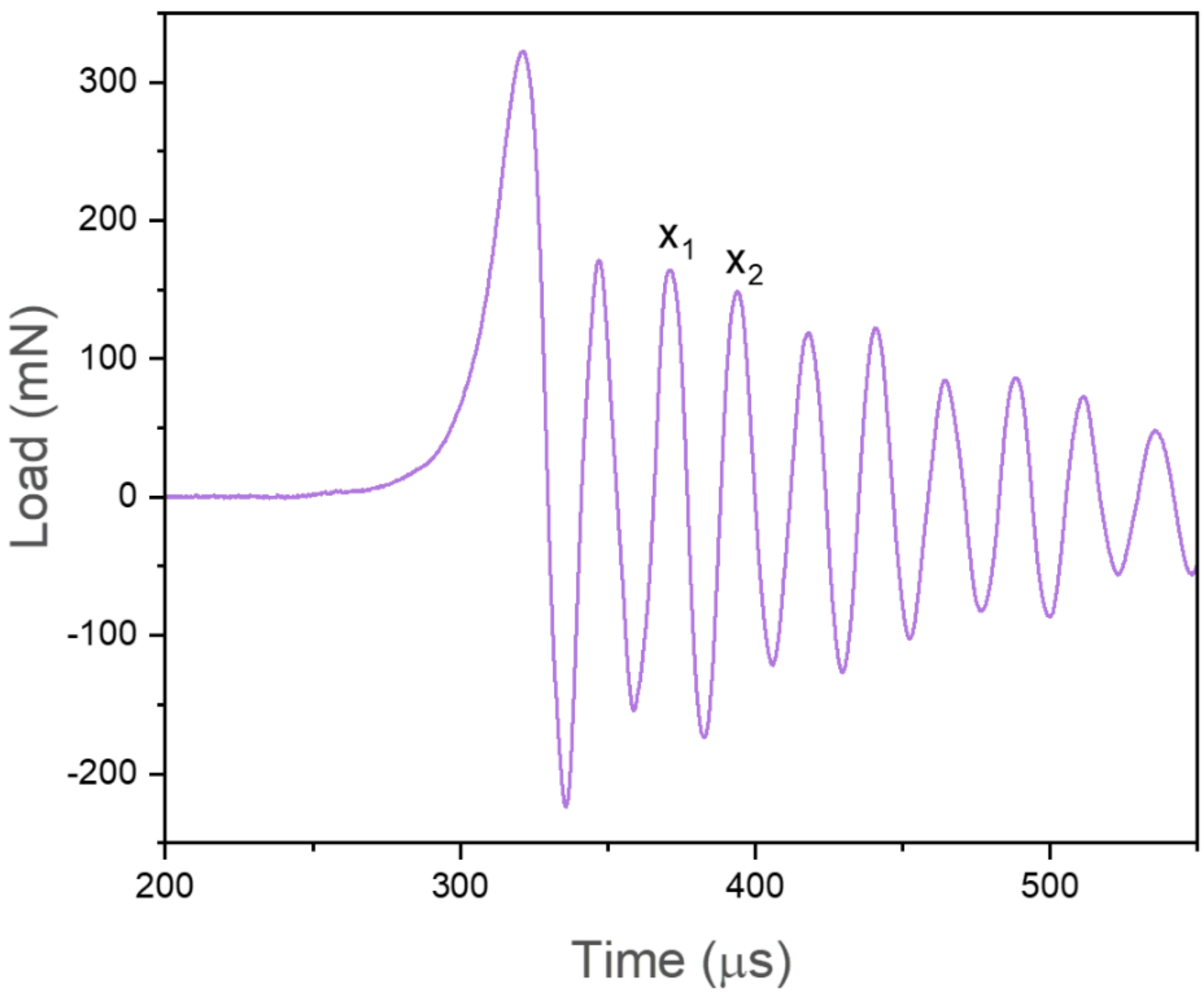

**Fig. S17.**

The free body oscillation of the piezoelectric load cell after an indentation experiment which was carried out at $5\times10^4$ $s^{-1}$ in molybdenum.

The concise flowchart for machine dynamics correction is as,

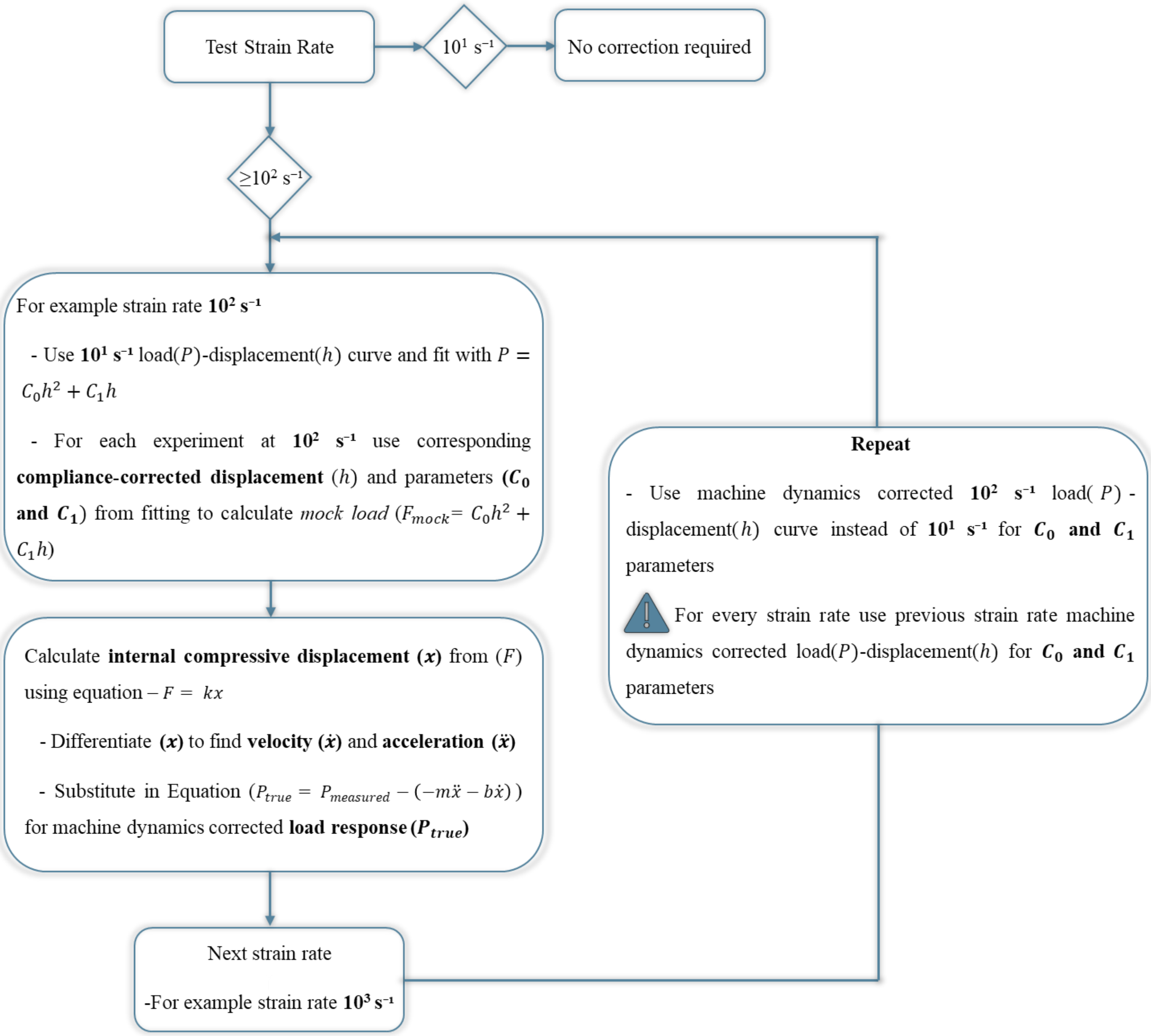

To evaluate the validity of these assumptions, an error analysis was performed on one of the load-displacement curves at the highest strain rate, $10^5$ s$^{-1}$ for molybdenum. The first step involves performing a quadratic fit on a machine dynamics-corrected load-displacement curve from a prior strain rate to obtain the $C_0$ and $C_1$ parameters as shown in fig. S18A. To examine the influence of these parameters, four different strain rates were selected: the machine dynamics-corrected load-displacement curves of $5\times10^4$ s$^{-1}$ corrected with $2.5\times10^4$ s$^{-1}$, $2.5\times10^4$ s$^{-1}$ corrected with $10^4$ s$^{-1}$, $10^4$ s$^{-1}$ corrected with $3\times10^3$ s$^{-1}$, and the load-displacement curve of $10^1$ s$^{-1}$. Additionally, $C_0$ and $C_1$ parameters were obtained by quadratic fitting of the uncorrected $10^5$ s$^{-1}$ curve. This approach was taken because, in the literature [6], a common simplification for correcting machine dynamics assumes that the measured load ($P_{measured}$) is same as the external force (F) used to calculate the internal displacement ($x$) of the secondary body. This study was aimed to understand the impact of this assumption. The $C_0$ and $C_1$ parameters, along with the compliance-corrected displacement (shown in the inset of fig. S18B), were used in the quadratic function to calculate the *mock load* ($F_{mock}$), as shown in fig. S18B. This *mock load* ($F_{mock}$) was used to determine the accurate compressive displacement ($x$) of the piezoelectric load cell using Equation (6). By differentiating ($x$) with respect to time, the velocity ($\dot{x}$) and acceleration ($\ddot{x}$) were derived and applying them to Equation (5) the true load response ($P_{true}$) corrected for machine dynamics could be obtained. Fig. S18C presents the machine dynamics-corrected load-displacement curves for $10^5$ s$^{-1}$, calculated using the different $C_0$ and $C_1$ parameters. Although fig. S18C shows minimal differences among the machine dynamics-corrected load-displacement curves, the load difference relative to the curve corrected using $C_0$ and $C_1$ parameters from $5\times10^4$ s$^{-1}$ are compared with the others and plotted in fig. S18D. One can observe that if one uses $C_0$ and $C_1$ parameters from $2.5\times10^4$ s$^{-1}$ the difference in load at 700 nm is less than ~ 1.3 mN and if it is $10^4$ s$^{-1}$ it is ~ 3.5mN and it becomes significantly higher only if $10^1$ s$^{-1}$ is used. Also, if the non-machine dynamic corrected $10^5$ s$^{-1}$ is used to obtain the $C_0$ and $C_1$ parameters the difference is in the order of ~ 1.3 mN. For perspective, this load variation leads to a hardness difference of approximately 0.15 GPa when using $C_0$ and $C_1$ parameters from $2.5\times10^4$ s$^{-1}$ and non-machine dynamic corrected $10^5$ s$^{-1}$. The difference increases to about 0.4 GPa and 0.9 GPa when using $C_0$ and $C_1$ parameters from $10^4$ s$^{-1}$ and $10^1$ s$^{-1}$, respectively. However, it is recommended to use the closest prior strain rate load-displacement response, corrected for machine dynamics, to determine the $C_0$ and $C_1$ parameters at high strain rates, thereby minimizing potential errors. An interesting observation is that while the loading portion exhibited changes in load, the peak load value remained unaffected by machine dynamics corrections, as shown in fig. S5B.

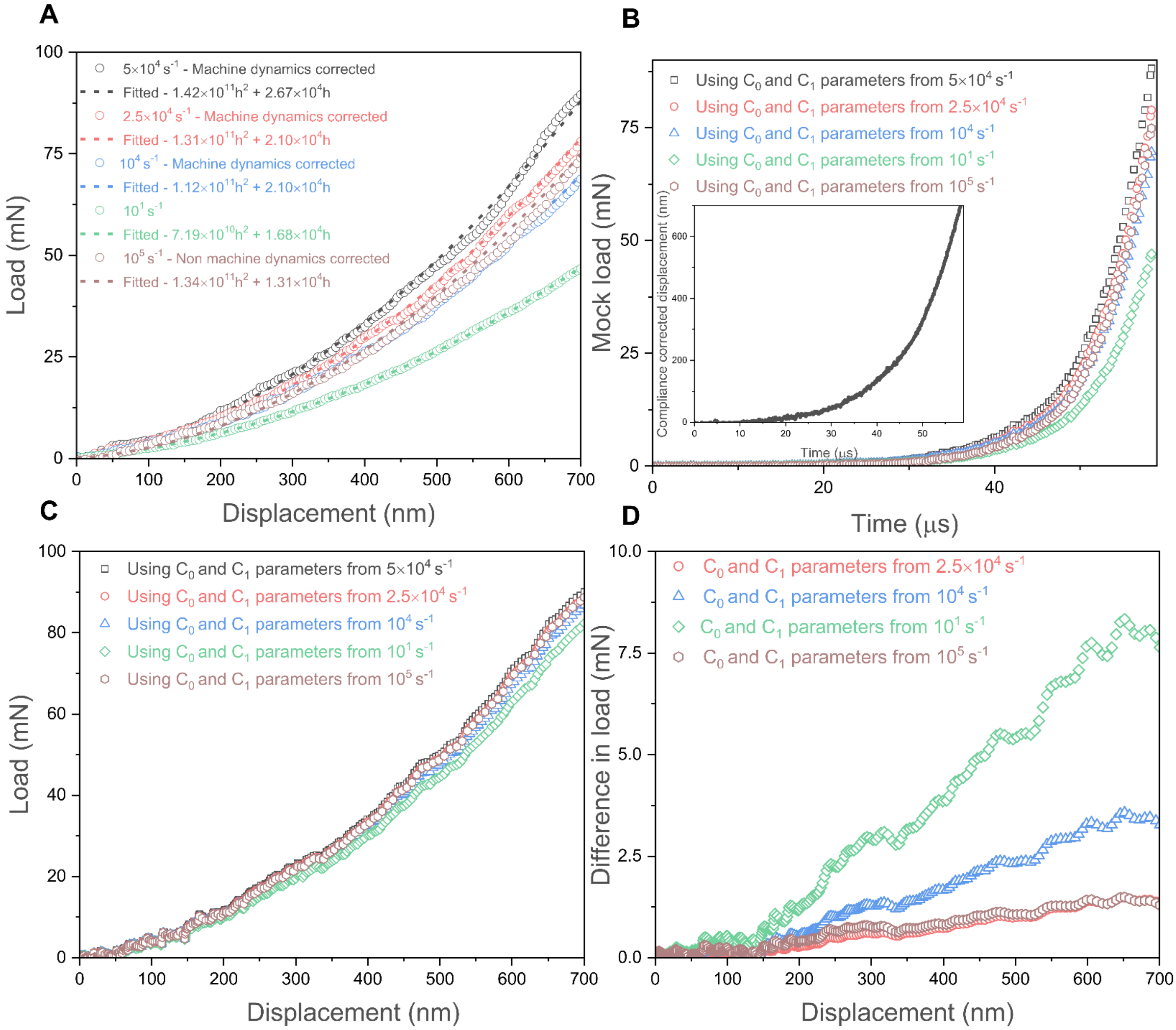

**Fig. S18.**

(**A**) The various load-displacement curves that were quadratically fitted to obtain the $C_0$ and $C_1$ parameters. (**B**) the *mock load* ($F_{mock}$) calculated using the parameters along with the compliance-corrected displacement (shown in the inset) in the quadratic function. (**C**) the machine dynamics-corrected load-displacement curves for $10^5$ s$^{-1}$, calculated using the $C_0$ and $C_1$ parameters obtained from the load-displacement curves in figure (**A**). (**D**) the load difference for all the curves in (c) relative to the curve corrected using $C_0$ and $C_1$ parameters from $5{\times}10^4$ s$^{-1}$.

***Section S4 - Nanoindentation experiments, micropillar compression testing, quasi-static reloading experiments, hardness measurement, and TEM, along with dislocation density measurement***

*Nanoindentation experiments*

All indentation experiments were performed using a Berkovich indenter tip from Synton MDP (Nidau, Switzerland). Tests were conducted across eight different strain rates, ranging from $10^1$ $s^{-1}$ to $10^5$ $s^{-1}$. Exponential voltage profiles were used for achieving constant indentation strain rate ($\dot{h}/h$). For all strain rates, the target indentation depth was consistently set to approximately 1.7 µm. It should be noted that this value represents the displacement, without accounting for system compliance corrections. At each strain rate, five experiments were carried out for statistically relevant results. The indenter was brought into contact with the sample using a pre-set load of 500 µN, with careful control to maintain a gap of approximately 10 nm between the tip and the sample before the start of each experiment. System compliance was measured using the multiple indentation method with a constant strain rate ($\dot{h}/h$) of $10^1$ $s^{-1}$ prior to conducting the high strain rate indentation experiments on each sample. All indentation experiments were conducted *in situ* within a Zeiss Gemini 500 (SEM) operating at 5 kV. For experiments involving fused silica, the electron beam was switched off during indentation. Indentation at a strain rate of $10^0$ $s^{-1}$ on fused silica was performed using the quasi-static setup[9] from Alemnis AG (Switzerland), while indentation at $5\times10^{-2}$ $s^{-1}$ was conducted using the G200 nanoindenter (KLA Corporation, USA) equipped with a standard Berkovich tip and employing the continuous stiffness method (CSM).

*Micropillar compression testing*

Compression tests were performed on lithographically fabricated fused silica micropillars with a diameter of 2.5 µm and height of 5.5 µm, at strain rates ranging from 1 $s^{-1}$ to 1750 $s^{-1}$ and on 5 µm and height of 14.5 µm, at strain rates ranging from 0.3 $s^{-1}$ to 1250 $s^{-1}$. The maximum compression achieved was 2 µm, at which point most pillars fractured. The load-displacement curves were converted into stress-strain data using the top cross-sectional area and pillar height. Since the majority of deformation occurred in the upper one-third of the pillar, this assumption for the top cross-sectional area in calculating the stress was considered valid. It was ensured that experiments were carried out at constant strain rates. The experiments were conducted *in situ* within a Zeiss Gemini 500 (SEM) operating at 5 kV. Prior to testing, a 30 µm diameter flat punch tip (from Synton MDP) was aligned with the top of the pillars, after which the electron beam was turned off for the compression experiments.

*Quasi-static reloading experiments*

The reloading experiments were conducted immediately following the high constant strain rate nanoindentation tests. After the initial high strain rate indentations, the indenter was automatically repositioned onto the pre-existing indent with a minimum auto-approach load of 500 μN. Subsequent reloading was performed in a displacement-controlled mode at a constant strain rate using the profile of $10^1$ $s^{-1}$. To ensure consistency, the initial system compliance corrected displacements during the high strain rate indentations were kept uniform across all experiments. The reloading hardness values were extracted from the load-displacement curves using the unloading stiffness based traditional Oliver-Pharr method [10]. It is important to note that, at the highest strain rates, the strain rates were not constant throughout the entire indentation depth (see fig. S1). Nonetheless, reloading experiments were performed on those indents.

*Hardness measurement*

In this study, hardness measurements were conducted using two complementary techniques. The projected area method, involved imaging the indents with a VK-X model confocal laser microscope from Keyence Corporation to accurately determine the contact area. The hardness (H) was then calculated using the relation:

$$H = \frac{P_{max}}{A_c} \qquad (14)$$

where $A_c$ is the measured contact area, and $P_{max}$ is the maximum load. The imaged indents were analyzed using Gwyddion 2.64 software. A polynomial background correction and baseline correction were done to the confocal images before extracting the contact area of the indents. Any pile-up or sink-in effects that could potentially affect the contact area measurement, and thereby influence the hardness calculation, are accounted for using the calibration method described by K. W. McElhaney *et al.*[11]

The MHP iterative method, is a modification of the traditional Oliver-Pharr equations, as proposed by B. Merle et al [12]. This approach assumes that the reduced elastic modulus $E_r$ remains constant between strain rates. If this assumption becomes invalid, no solution is obtained. Given reliable load-displacement curves and tip area function parameters $\sum_{i=0}^{n} m_i$, hardness as a function of displacement can be determined. The brief description along with equations and MATLAB script for the MHP iterative method are as follows,

*MHP Iterative method*

This method is a modification of the traditional Oliver-Pharr equations, as proposed by B. Merle et al[12]. This approach assumes that the reduced elastic modulus ($E_r$) remains constant between strain rates. From standard Oliver–Pharr equations [13] the contact stiffness (S) and contact depth ($h_c$) is given by the following equation,

$$S = \frac{2E_r\beta\sqrt{A_c}}{\sqrt{\pi}} \qquad (15)$$

$$h_c = h - \varepsilon\frac{P}{S} \qquad (16)$$

Where, β and ε are constants dependent on the indenter geometry, with values close to 1.0 and 0.75, respectively. And the contact area ($A_c$) is given by,

$$A_c = \sum_{i=0}^{n} m_i h_c^{2^{1-i}} \quad (17)$$

In this study, three contact area parameters ($m_0$ = 25.03886, $m_1$ = -5.53811E-7, and $m_2$ = 3.25942E-10) were utilized to model the tip shape accurately. These parameters were derived using continuous stiffness measurement (CSM) on fused silica at a strain rate of $10^{-2}$ $s^{-1}$ with an oscillation frequency of 10 Hz and an amplitude of 20 nm, performed using a quasi-static testing setup, detailed elsewhere [9].

Using equations (15-17), the equation was re-written to the following form to eliminate the contact stiffness (S),

$$\left\{\left(\sum_{i=0}^{n} m_i h_c^{2^{1-i}}\right) \times (h_c - h)^2\right\} - \frac{\pi\varepsilon^2 P^2}{4\beta^2 {E_r}^2} = 0 \qquad (18)$$

Using an iterative technique, a solution for $h_c$ less than the indentation depth (h) was identified. From the $h_c$, $A_c$ was calculated using equation (17). Then using the following equation hardness (*H*) was calculated,

$$H = \frac{P}{A_c} \qquad (19)$$

where $A_c$ is the measured contact area, and *P* is the applied load. Using this method, hardness could be calculated continuously over the entire displacement.

The MATLAB script for the iterative method is as follows,

```
% The input file should be in .CSV format, with the 1st row as time(s), 2nd row
% as displacement(m) and 3rd row as load(N)
clc;
clear all;
format shortE;
M = readtable('inputfilename.csv', 'DecimalSeparator',',');
M = M{:,:};
```

```
Ti = M(:,1);
Disp = M(:,2);
Load = M(:,3);
Di = smoothdata(Disp,"sgolay",50);
Lo = smoothdata(Load,"sgolay",50);
Er =; % enter reduced elastic modulus here
m0 =;% enter contact area parameters
m1=;% enter contact area parameters
m2=;% enter contact area parameters
L=length(Di);
Z=0;
j1=0;
for i1 = 1:1:L
    Z = Z+1;
    d = Di(i1,1);
    l = Lo(i1,1);
    in = d/100000;
    j=0;
    for x = 0:in:d
    j=j+1;
    h(j,1)= x;
    y(j,1)= (((m0*(x^2))+ (x*m1)+ ((x^0.5)*m2))*((d-x)^2)) -
((3.14*0.5625*(l^2))/(4.276624*(Er^2)));
    end
    zci = @(v) find(diff(sign(v))); % Returns Approximate Zero-Crossing Indices Of
Argument Vector
    zx = zci(y);
    M = max (zx);
    Sol = h(M,1);
    hc = polyval(Sol,1);
    Ac = ((m0*(hc^2))+ (hc*m1)+ ((hc^0.5)*m2));
    Hard = l/Ac;
    j1=j1+1;
    H(j1,1) = d;
    H(j1,2) = Hard;
    H(j1,3) = hc;
    H(j1,4) = Ac;
end
plot(H(:,1),H(:,2))
head={'Displacement [m]' 'Hardness [GPa]' 'hc [m]' 'Ac [m2]' };
F=[head;num2cell(H)];
writecell(F,'filenametosave.csv','Delimiter','\t')
```

*TEM and dislocation density calculations*

Lamellae for TEM analysis were prepared using a Thermofischer Scientific Scios 2 focused ion beam-scanning electron microscope (FIB-SEM) equipped with a Ga-ion beam. Lamellae were lifted out from the central regions of the indents (along the median axis of the indent triangle). To ensure consistent cross-sectional orientations, all lamella were prepared in the same direction relative to the indents. The lift-out process was performed at an accelerating voltage of 30 kV, followed by final cleaning at 5 kV and 48 pA. Annular bright field scanning transmission electron microscopy (ABF-STEM), selected area electron diffraction (SAED), and convergent beam electron diffraction (CBED) was performed on a Thermo Scientific Titan

Themis at 300 keV with a convergence angle of 4 mrad. Correlative images were simulated using the program Dr. Probe [14] to find the thickness of the lamella.

Dislocation density measurements were performed using ABF-STEM images. The detailed procedure for extracting dislocation density from a lamella indented at $10^1$ s$^{-1}$ is provided, and a similar methodology was used for lamellae indented at $3\times10^3$ s$^{-1}$ and $5\times10^4$ s$^{-1}$. Due to the extremely high dislocation density, as qualitatively evident from the ABF-STEM images in Fig. 4 of the main paper, the traditional line intercept method was not employed. Instead, an alternate method was adopted that calculates the area fraction of pixels covered by dislocations in a sample of given thickness 't'. This approach is detailed by J. Gallet *et al*.[15]. According, to this technique the dislocation density is calculated using the following formulation,

$$\rho = \frac{N^{dislo}}{N^{tot} \times t \times E^{app}} \qquad (20)$$

where, $N^{dislo}$ is the total number of dislocation pixels, $N^{tot}$ is the total number of pixels in the analysed image and $E^{app}$ refers to the apparent dislocation width (in nm) measured from the image at the same magnification used for dislocation density measurements.

Fig. S19A shows the ABF-STEM image of the $10^1$ s$^{-1}$ indent, focusing on a region approximately 1.5 μm below the indent tip. For dislocation density measurement, a 1.5 × 1.5 μm² area within this region was selected, as shown in fig. S19B. The region 1.5 μm below the indent was chosen because, at higher strain rates, the dislocation density is so high that quantifying it reliably becomes challenging. To maintain consistency across all analyzed indents, the same depth below the indent was selected. After which a threshold was applied in the grayscale value between 0-100 and the ratio of $\frac{N^{dislo}}{N^{tot}}$ was calculated as shown in fig. S19C. The grayscale value was kept constant across all analysed strain rates. Similarly, $E^{app}$ was determined by averaging the apparent dislocation width across several dislocations at the same magnification used for dislocation density measurements, as shown in fig. S19D. By substituting all the measured values, along with the lamella thickness determined from the CBED analysis, into equation (20), the dislocation density was calculated.

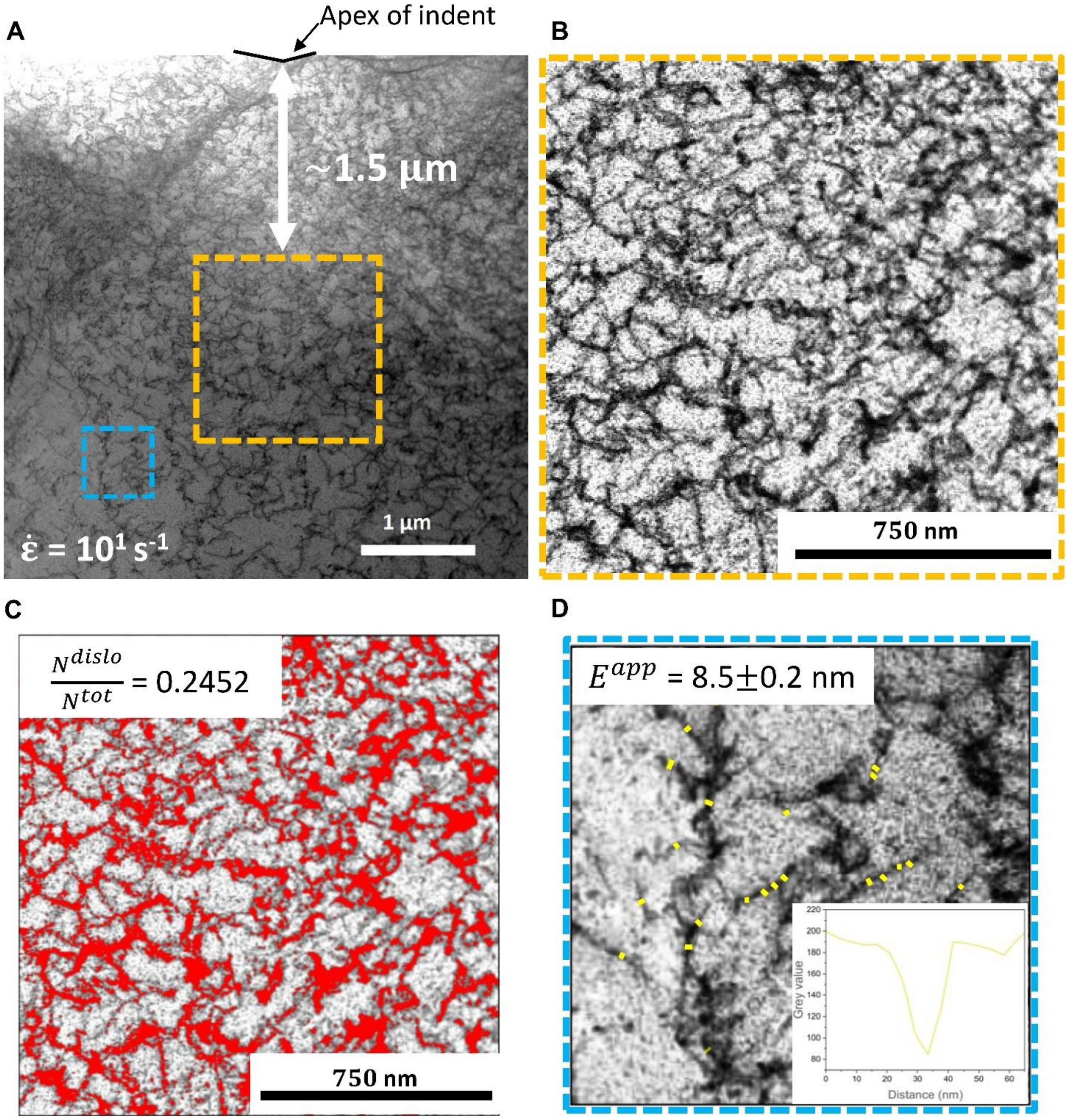

**Fig. S19.**

(**A**) the ABF-STEM image of the $10^1$ s$^{-1}$ (**B**) a region of 1.5 × 1.5 µm$^2$ area at a distance of 1.5 µm from the indent tip (marked with black lines) (**C**) the same region as in (b) with a grayscale threshold applied (0–100) to determine the $\frac{N^{dislo}}{N^{tot}}$ fraction and (**D**) a selected area from (**A**) used to measure the apparent dislocation width ($E^{app}$), where the yellow lines indicate the dislocations on which the width measurements were made using a line profile as shown in inset.